\documentclass{aa}  
\usepackage{graphicx,url}
\usepackage[normalem]{ulem}
\usepackage{mathrsfs,amssymb,amsmath}
\usepackage[usenames]{color}
\usepackage[varg]{txfonts}
\usepackage{booktabs}
\usepackage{multirow}

\usepackage{bm}

\usepackage[breaklinks, colorlinks, citecolor=blue]{hyperref}
\newcommand{\be}{\begin{equation}}
\newcommand{\ee}{\end{equation}}

\newcommand{\ud}{{\rm d}}

\newcommand{\vecbperp}{${\bf B}_\perp$}
\newcommand{\bperp}{$B_\perp$}
\newcommand{\bperpmath}{B_\perp}

\newcommand{\Vr}{\mathbf{r}}

\begin{document}

\title{Spectral index of synchrotron emission:\\insights from the diffuse and magnetised interstellar medium} 

 \author{Marco Padovani
          \inst{1}
          \and
          Andrea Bracco\inst{2}
          \and
          Vibor Jeli\'c\inst{2}
          \and
          Daniele Galli\inst{1}
          \and
          Elena Bellomi\inst{3}
          }

   \institute{INAF--Osservatorio Astrofisico di Arcetri, 
              Largo E. Fermi 5, 50125 Firenze, Italy\\
              \email{marco.padovani@inaf.it}
              \and
              Ru{\dj}er Bo\v{s}kovi\'c Institute, Bijeni\v{c}ka cesta 54, 
              10000 Zagreb, Croatia 
            \and
Observatoire de Paris, LERMA, Sorbonne Universit\'e, CNRS, Universit\'e PSL, 75005 Paris, France\\
             }


 
  \abstract
{The interpretation of Galactic synchrotron
observations is complicated 
by the degeneracy between the strength of the magnetic field 
perpendicular to the line of sight (LOS), \bperp, 
and the cosmic-ray electron (CRe) spectrum. Depending
on the observing frequency, an energy-independent 
spectral energy slope $s$ 
for the CRe spectrum is usually assumed: $s=-2$ at frequencies
below $\simeq$400 MHz and $s=-3$ at higher frequencies.}
{
Motivated by the high angular and spectral resolution of current facilities such as the LOw
Frequency ARray (LOFAR) and future telescopes such as the Square Kilometre Array (SKA), 
we aim to understand the consequences of taking 
into account the energy-dependent CRe spectral energy slope 
on the analysis of the spatial variations of the brightness temperature spectral index, $\beta$, 
and on the
estimate of the average value of \bperp\ along the LOS.
}
  {We illustrate analytically and numerically the impact that different realisations of the CRe spectrum   
   have on the interpretation of the spatial variation of
   $\beta$. We use two snapshots from 3D magnetohydrodynamic simulations
   as input for the magnetic field, 
   with median magnetic field strength of $\simeq4$ and $\simeq20~\mu$G, to study
   the variation of $\beta$ over a wide range of frequencies ($\simeq0.1-10$~GHz).}
 {We find that the common assumption of an 
 energy-independent $s$ is valid only in special cases.
  We show that 
  for typical magnetic field strengths of the diffuse ISM ($\simeq$2$-$20~$\mu$G),
  at frequencies of 0.1$-$10~GHz, 
  the electrons that are mainly responsible for the synchrotron emission have energies in the 
  range $\simeq$100~MeV$-$50~GeV. This is the energy range where the spectral slope, $s$, of CRe has its greatest variation.
   We also show that the polarisation fraction can be much smaller than the maximum value of 
   $\simeq 70\%$ 
   because the orientation of \vecbperp\ varies across the telescope's beam and along the LOS.
   Finally, we present a look-up plot that can be used
   to estimate the average value of $B_\perp$ 
   along the LOS from a set of values of  $\beta$ 
   measured at different frequencies, for a given CRe spectrum.
   }
   {In order to interpret the spatial variations of $\beta$ observed  from centimetre to metre wavelengths across the Galaxy, the 
   energy-dependent slope of the Galactic CRe spectrum in
   the energy range $\simeq$100~MeV$-$50~GeV 
   must be taken into account.
   }


   \maketitle
%
\section{Introduction}

Studies of diffuse synchrotron emission and its polarisation play
a key role in constraining properties of the magnetic fields and
of the interstellar medium (ISM) in the Milky Way, especially the
cosmic-ray electron (CRe) energy spectrum. They are also relevant for
measurements of the cosmic microwave background radiation at high
radio frequencies \citep[e.g.,][and references therein]{Planck2018IV}
and the cosmological 21~cm radiation from Cosmic Dawn and Epoch of
Reionisation (EoR) at low radio frequencies \citep[e.g.,][]{Bowman2018,
Gehlot2019, Mertens2020, Trott2020}. Galactic synchrotron emission
is one of the main foreground contaminants in these cosmological
experiments and its emission dominates the radio sky at frequencies
below $10~{\rm GHz}$. It is therefore of great importance to
understand in detail the spectral and spatial variations of Galactic
synchrotron emission in order to successfully mitigate it in
cosmological observations (for more details see a review by
\citealt{ChapmanJelic19}). 
    
The spectrum of Galactic synchrotron emission
is usually expressed in terms
of the brightness temperature $T_\nu$ (defined in Sect.~\ref{basiceqs})
characterised by a frequency-dependent spectral index 
$\beta(\nu)=d\log T_{\nu}/d\log\nu$.
Spatial variations of $\beta$ in a given region of the sky reflect spatial variations of the CRe and magnetic field properties in the ISM along the line of sight (LOS) across that region. 
Full sky maps of Galactic synchrotron emission 
clearly show spatial variations of $\beta$
already at the angular resolution of 
$\simeq5^\circ$ \citep{Guzman+2011}. 
Facilities providing at least three times this angular resolution such as 
the LOw Frequency ARray, LOFAR
\citep{vanHaarlem2013},
and in the
near future the Square Kilometre Array, 
SKA \citep{Dewdney+2009}, will be 
able to investigate even finer variations of $\beta$.

In terms of frequency, the observed synchrotron spectrum  is flatter at low radio frequencies than at high radio frequencies \citep{Roger+1999,Guzman+2011}. Typical values of 
$\beta$ estimated from
observations
at mid and high Galactic latitudes are $-2.59 < \beta < -2.54\pm0.01$ between 50 and~100 MHz \citep{Mozdzen2019} and $-2.62 \pm 0.02<\beta<-2.60$ between 90 and~190 MHz \citep{Mozdzen2017}, as measured recently by the Experiment to Detect the Global EoR Signature (EDGES). 
In contrast, in the
frequency range 1.4$-$7.5~GHz,
$\beta$ is $-2.81\pm0.16$
\citep{Platania+1998}.
This difference in the spectral index at low and high radio frequencies is related to ageing of the CRe energy spectrum
(hereafter simply
``CRe spectrum''), $j_e(E)$.

As CRe propagate through the ISM, they lose energy by a number of energy-loss mechanisms that involve interactions with matter, magnetic fields, and radiation \citep{Longair2011}. These processes deplete the population of relativistic electrons and change their original energy (injection) spectrum. 
The knowledge of the Galactic CRe spectrum is a challenging subject that has seen significant advances only 
in recent years. 
A number of relevant results have been achieved at high energies (above $\simeq10$~GeV) thanks to the detections of the {\em Fermi} Large Area Telescope \citep[{\em Fermi} LAT;][]{Ackermann+2010}, the balloon-borne Pamela experiment \citep{Adriani+2011}, and the Alpha Magnetic Spectrometer (AMS-02) on board of the International Space Station \citep{Aguilar+2014}. Only recently the two Voyager probes crossed the heliopause, overcoming the problem of solar modulation, and constraining $j_e(E)$
down to $E\simeq 3$~MeV. \citep{Cummings+2016,Stone+2019}.
Yet, the origin and propagation of CRe partly remain unresolved due to the degeneracy of a number of parameters and uncertainty about the role of reacceleration, convection, and on the diffusion coefficient (see, e.g., \citealt{Strong+2007} and \citealt{Grenier+2015} for comprehensive reviews on this topic).

The uncertainties on the 
Galactic CRe spectrum limit the interpretation of Galactic synchrotron emission.
Depending on the frequency of observation, it is usually assumed that the spectrum of the electrons contributing to the emission can be characterised by
a single energy slope. 
In this paper we will show that this assumption results in an oversimplification 
and needs to be replaced by a more accurate modelling when interpreting observations 
from centimetre to
metre wavelengths. 
This is especially the case for the recent LOFAR \citep{vanHaarlem2013} polarimetric observations \citep[e.g.][]{Jelic+2015, vanEck2017}, where observed polarised structures were possibly associated with 
synchrotron radiation from
neutral clouds. As suggested by \citet{vanEck2017} and further supported 
by \citet{Bracco+2020}, the observed polarised synchrotron emission might be originating from low column density clouds of interstellar gas along the sight line composed of a mixture of warm and cold neutral hydrogen media, referred to as WNM and CNM, respectively.

In the light of these recent results, we partly focus on the effects of the energy-dependent CRe spectral energy slope at a few hundred MHz. However, we also highlight the impact of an 
energy-dependent energy slope
at higher frequencies.
This paper is organised as follows. In Sect.~\ref{sec:basics} we introduce the theory of synchrotron emission and illustrate the effects on the brightness temperature spectral index depending on the parameterisation of the 
CRe spectrum;  in Sect.~\ref{sec:modelsynem} we apply 
the above results 
first to a cloud modelled as a uniform slab, then to the diffuse, multiphase ISM with the help of 3D magnetohydrodynamic (MHD) simulations, that we then use in Sect.~\ref{sec:results} to compute brightness temperature maps, the spectral index, and the polarisation fraction. In Sect.~\ref{sec:discussion} 
we discuss the effect of the angular resolution of the observations and show a procedure for predicting the average strength of 
the magnetic field perpendicular to the LOS, \bperp, 
once the CRe spectrum is set. 
In Sect.~\ref{sec:summary} we summarise the main findings.

\section{Fundamentals of synchrotron emission}
\label{sec:basics}

Above $\simeq 10$~GeV, the CRe spectrum,
(i.e., the number of electrons per unit energy, time, area, and solid angle),
is approximately a 
power law in energy, $j_e(E)\propto E^s$. The spectral energy slope
(hereafter 
simply ``spectral slope''), $s(E)=d \log j_e/d\log E$, has 
been measured in the solar neighbourhood by 
several probes: 
{\em Fermi}
LAT established a spectral slope
$s=-3.08\pm0.05$ in the energy
range 7~GeV--1~TeV \citep{Ackermann+2010},
the Pamela experiment found 
$s=-3.18\pm 0.05$ above the energy region influenced by the solar wind ($>30$~GeV; \citealt{Adriani+2011}),
and AMS-02 
determined $s=-3.28\pm 0.03$ in the 
energy range 19.0--31.8~GeV and 
$s=-3.15\pm 0.04$ in the range
83.4--290~GeV \citep{Aguilar+2014}.
At low energies the spectral slope
measured by the two Voyager probes in the energy range $\simeq$3$-$40~MeV is $-1.30\pm 0.05$ 
\citep{Cummings+2016,Stone+2019}. Thus, at energies below $\simeq$10~GeV, 
the spectral slope is energy-dependent. As we will see, this has 
significant consequences on the spectrum of the synchrotron emission observed at frequencies of 
hundreds of MHz up to tens of GHz.

\begin{figure}[!h]
\begin{center}
\resizebox{1\hsize}{!}{\includegraphics{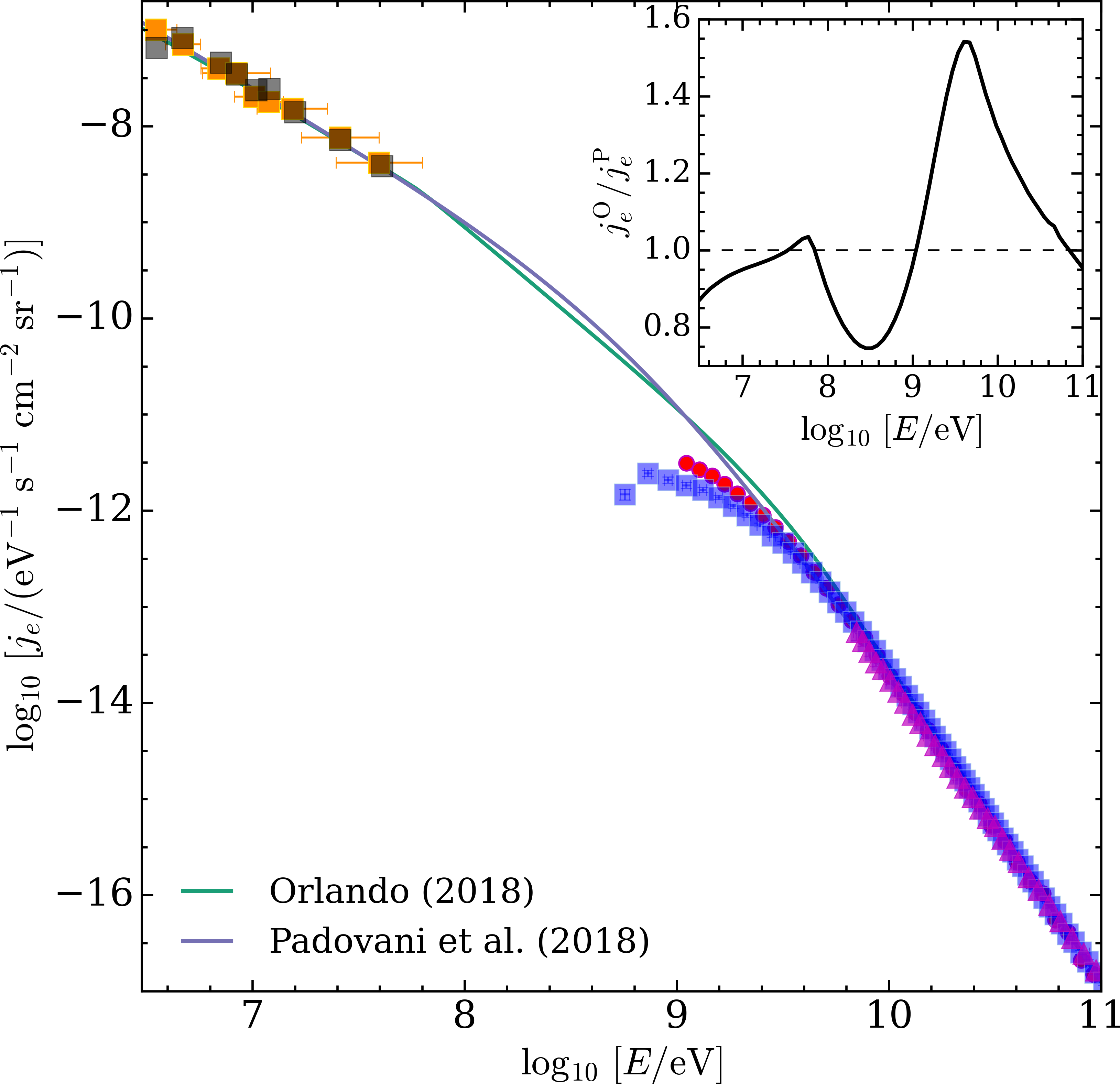}}
\caption{CRe spectrum according to \citet{Orlando2018} (green line)
and \citet{Padovani+2018a} (violet line).
The inset shows the ratio of the spectra
(O=Orlando, P=Padovani; solid black line).
Data: 
Voyager~1 \citep[orange squares;][]{Cummings+2016};
Voyager~2 \citep[grey squares;][]{Stone+2019};
{\em Fermi} LAT \citep[magenta triangles;][]{Ackermann+2010};
Pamela \citep[red circles;][]{Adriani+2011};
AMS-02 \citep[blue squares;][]{Aguilar+2014}.
 }
\label{electronspectra}
\end{center}
\end{figure}

Generally, models and simulations developed for the interpretation of Galactic synchrotron emission assume 
that CRe contributing to 
the emission above and below 408~MHz have
$s=-3$ and $s=-2$, respectively \citep[see, e.g.,][]{Sun+2008,Waelkens+2009,Reissl+2019,Wang+2020}. This simplification is usually made to avoid time-consuming calculations, and is based on observations supporting a flatter spectrum below 408~MHz
\citep[see, e.g.,][]{ReichReich88a,ReichReich88b,Roger+1999}.
In Sect.~\ref{ssec:srange} we show that this assumption
turns out to be inaccurate
both at low and high frequencies, leading to 
a misinterpretation of the synchrotron
observations, in particular of the spatial variations of the
spectral index $\beta$.
To support this claim we consider the two realisations of the
local
CRe spectrum by \citet{Orlando2018} and 
\citet{Padovani+2018a} shown in Fig.~\ref{electronspectra}. The former is based on multifrequency 
observations, from radio to $\gamma$ rays, and Voyager~1 measurements
through propagation models, and is representative of intermediate Galactic latitudes ($10^{\circ}<|b|<20^{\circ}$) that include most of the local radio synchrotron emission within $\sim 1$~kpc around the Sun;
the latter is given by an analytical four-parameter fitting formula that 
reproduces exactly the power-law behaviour measured at low
and high energies \citep[see also][]{Ivlev+2015}. 
As shown by the inset in Fig.~\ref{electronspectra}, the 
spectra by \citet{Orlando2018} and that of  
\citet{Padovani+2018a} differ by less than $\sim 25$\% over the range
of energies of interest here, below $E\simeq 50$~GeV.
As we will see in Sect.~\ref{ssec:betaO18P18}, synchrotron observations 
can also constrain 
these two parameterisations.

\begin{figure}[!h]
\begin{center}
\resizebox{.975\hsize}{!}{\includegraphics{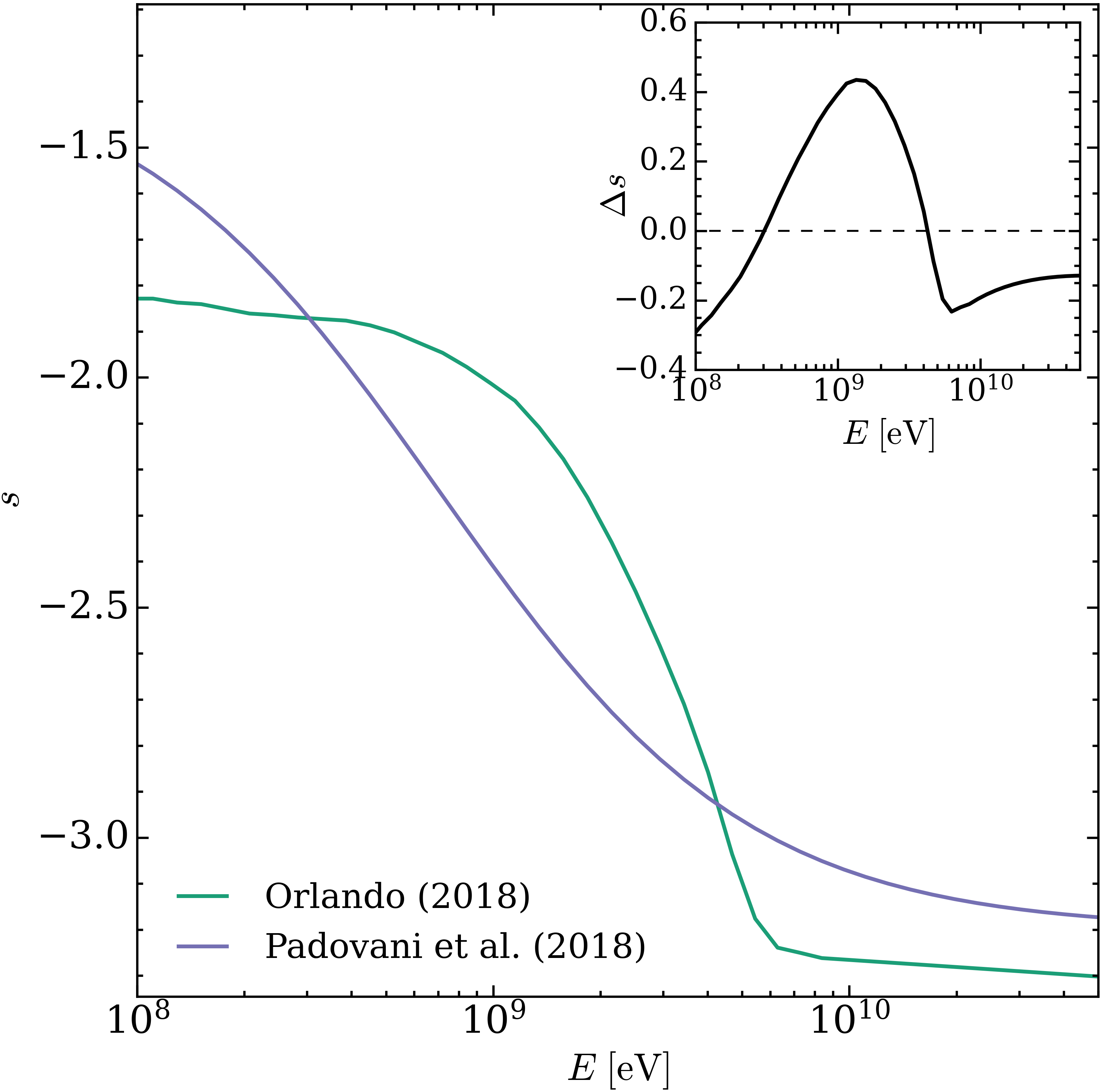}}
\caption{Spectral slope versus energy of the CRe
spectra by \citet{Orlando2018} (green line)
and \citet{Padovani+2018a} (violet line)
in the energy range relevant for our study (100~MeV$-$50~GeV). 
The inset shows the the difference $\Delta s$ between the spectral slope of the two spectra.
}
\label{energyslope}
\end{center}
\end{figure}

CRe at a given energy emit over a broad range of frequencies and, conversely,
the synchrotron emission observed at a given frequency comes from a broad range of CRe
energies. 
Before focusing on the frequency range of LOFAR observations (115--189~MHz; see Sect.~\ref{sec:results}), we consider 
frequency ranges characteristic of all-sky Galactic radio surveys both at low 
(45--408~MHz; \citealt{Guzman+2011}) and high frequencies (1--10~GHz; \citealt{Platania+1998}). 
For $\bperpmath\simeq 2$ to $20~\mu$G, as
expected in the diffuse ISM 
\citep[e.g.,][]{Heiles2005a,Beck2015,Ferriere2020},
CRe that account for nearly all of the observed synchrotron emission 
have energies ranging from $\simeq 100$~MeV to 50~GeV (see Sect.~\ref{ssec:srange})%
\footnote{We note that CRe in the energy range 100~MeV--50~GeV are
affected by energy losses such as bremsstrahlung only after crossing column 
densities $\gtrsim 10^{25}$~cm$^{-2}$ \citep[see, e.g.,][]{Padovani+2009,Padovani+2018a},
much larger than those typical of the diffuse medium. Therefore we will compute
synchrotron emissivities without accounting for any attenuation of the 
CRe spectrum.}.
In this energy range 
the spectral slope $s$ has a large variation with energy:
between $-1.8$ and $-3.3$ and between $-1.5$ and $-3.2$,
for the spectra modelled by \citet{Orlando2018} and \citet{Padovani+2018a},
respectively (see Fig.~\ref{energyslope}).

\subsection{Basic equations}
\label{basiceqs}

Here we summarise the basic equations for the calculation of the synchrotron brightness temperature, 
$T_\nu$ \citep[see, e.g.][for details]{GinzburgSyrovatskii1965}.
At any given position ${\bf r}$ in a cloud, the specific emissivity\footnote{The specific emissivity has units of
power per unit volume, frequency, and solid angle.} 
can be split into two components linearly polarised
along and across the component of the magnetic field perpendicular to the LOS, \vecbperp${\bf (r)}$,
\begin{eqnarray}\label{epsnu}
\varepsilon_{\nu,\|}(\Vr) &=& \int_{m_{e}c^{2}}^{\infty}\frac{j_{e}(E)}{v_{e}(E)}P_{\nu,\|}^{\rm em}(E,\Vr)\,\ud E,\\\nonumber
\varepsilon_{\nu,\perp}(\Vr) &=& \int_{m_{e}c^{2}}^{\infty}\frac{j_{e}(E)}{v_{e}(E)}P_{\nu,\perp}^{\rm em}(E,\Vr)\,\ud E\,,
\end{eqnarray}
where 
\begin{eqnarray}
\label{power}
P_{\nu,\|}^{\rm em}(E,\Vr)&=&\frac{\sqrt{3}e^{3}}{2m_{e}c^{2}} B_{\perp}(\Vr) [F(x)-G(x)],\\\nonumber
P_{\nu,\perp}^{\rm em}(E,\Vr)&=&\frac{\sqrt{3}e^{3}}{2m_{e}c^{2}} B_{\perp}(\Vr) [F(x)+G(x)]
\end{eqnarray}
are the power per unit frequency emitted by an electron of energy $E$ at frequency $\nu$ for the two polarisations.
Here, $B_\perp=|{\bf B}_\perp|$, $v_e$ is the electron velocity,
$x=\nu/\nu_c$, and $\nu_c$ is the critical frequency given by
\be\label{nuc}
\nu_c[B_{\perp}(\Vr),E]=\frac{3eB_{\perp}(\Vr)}{4\pi m_{e}c}\left(\frac{E}{m_e c^2}\right)^{2}\,.
\ee
The functions $F(x)$ and $G(x)$ are defined by
\be
F(x)=x\int_{x}^{\infty}K_{5/3}(\xi)\ud\xi
\ee
and
\be
G(x)=x K_{2/3}(x)\,,
\ee
where $K_{5/3}$ and $K_{2/3}$ are the modified Bessel functions of order
5/3 and 2/3, respectively.
The corresponding Stokes $Q_\nu$ and $U_\nu$ specific emissivities are
\be
\varepsilon_{\nu,Q}(\Vr)=[\varepsilon_{\nu,\perp}(\Vr)-\varepsilon_{\nu,\|}(\Vr)]\cos[2\varphi(\Vr)]
\ee
and
\be
\varepsilon_{\nu,U}(\Vr)=[\varepsilon_{\nu,\perp}(\Vr)-\varepsilon_{\nu,\|}(\Vr)]\sin[2\varphi(\Vr)]\,,
\ee
where $\varphi({\bf r})$ is the local polarisation angle counted positively clockwise. 
The orientation of \vecbperp\ rotated by $\pm90^{\circ}$ gives the local polarisation angle (modulo
$180^{\circ}$).
The emissivities are integrated along the LOS  
to obtain the specific intensity
(brightness) for each polarisation, $I_{\nu,\|}$ and $I_{\nu,\perp}$,
and the Stokes parameters $Q_\nu$ and $U_\nu$. 
We also compute the polarisation fraction 
\be\label{polfraceq}
p = \frac{P_\nu}{I_\nu} = \frac{\sqrt{Q_\nu^2+U_\nu^2}}{I_\nu}\,,
\ee
where $I_\nu=I_{\nu,\|}+I_{\nu,\perp}$. 
Finally, the flux density, $S_\nu$, obtained from the convolution
of the specific intensity integrated along the LOS with the telescope beam,
is converted into brightness temperature by the relation
\be\label{TnuSnu}
T_\nu = 10^{-23} \frac{S_\nu c^2}{2k_{\rm B}\Omega\nu^2}~{\rm K}\,,
\ee
where $\Omega=\pi\theta_b^2/(4\ln 2)$, $\theta_b$ is the 
full width of the beam at half its maximum intensity,
and $k_{\rm B}$ is the Boltzmann constant.
Here, $S_\nu$ is in Jy/beam and the other quantities are in cgs units. 
If the CRe spectrum is a single 
power law, $j_e(E)\propto E^{-s}$,
then $S_\nu\propto \nu^\alpha$, with $\alpha=(s+1)/2$, and $T_\nu \propto \nu^\beta$, with $\beta=(s-3)/2$ \citep{GinzburgSyrovatskii1965}.

\subsection{Contributions to synchrotron emission from 100~MHz to 10~GHz}
\label{ssec:srange}

\begin{figure*}[!h]
\begin{center}
\resizebox{1\hsize}{!}{\includegraphics{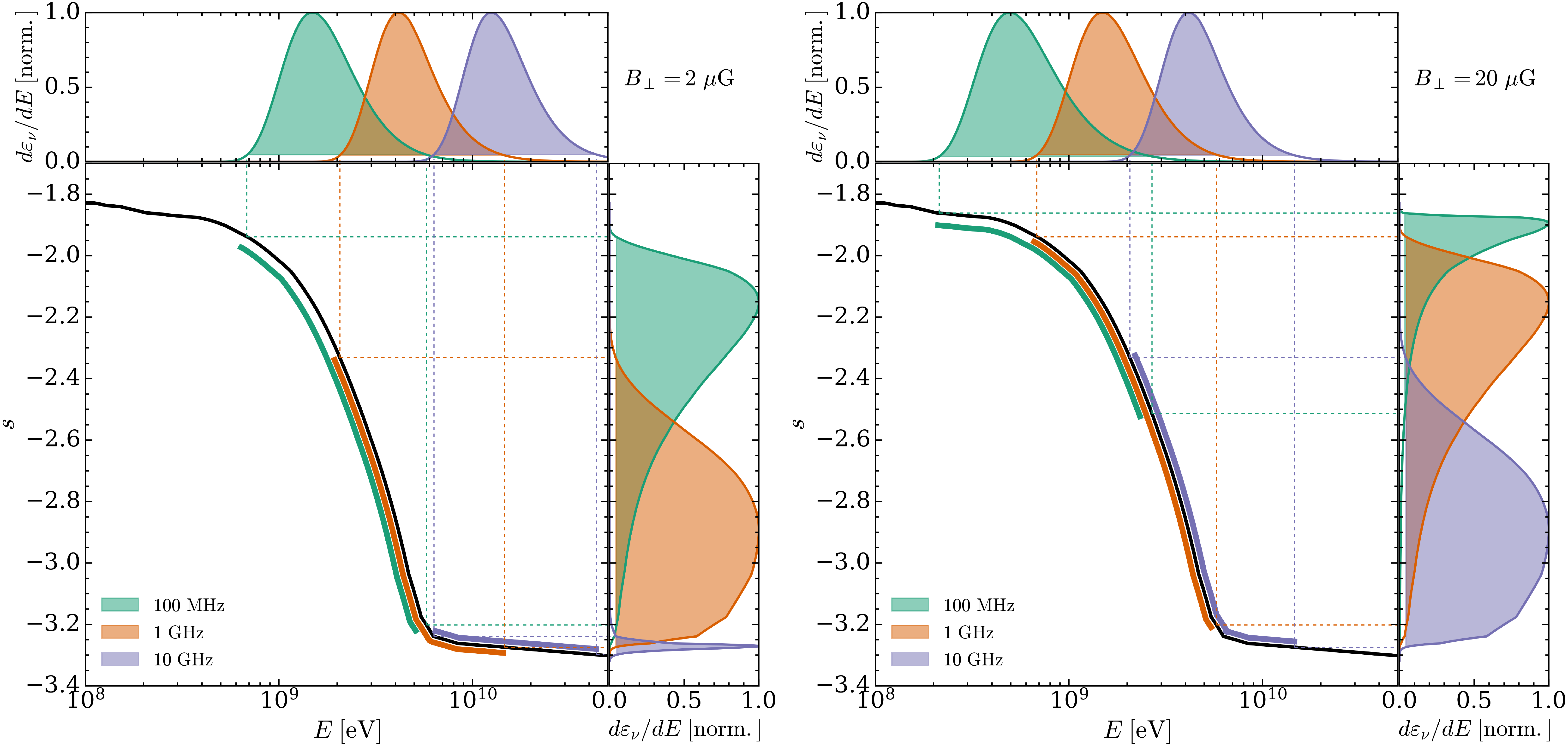}}
\caption{Left panel. 
Upper plot: integrand of the specific emissivity, $d\varepsilon_{\nu}/d E$, as a function of the energy $E$, normalised to its maximum value and summed over the 
two polarisations (Eqs.~(\ref{epsnu})), 
computed at 100~MHz, 1~GHz, and 10~GHz (green, orange, and violet line, 
respectively), for $\bperpmath=2~\mu$G. 
The coloured area below each curve corresponds to the 95\% of the total emissivity.
Right plot: same as the upper plot, as a function of the CRe spectral slope $s$.
Main plot: Spectral slope as a function of the energy for the CRe spectrum by \citet{Orlando2018} (black solid line).
Coloured lines show the range of energies (and the corresponding spectral slopes) contributing to 95\%
of the specific emissivity. 
Right panel: same as left panel, but for $\bperpmath=20~\mu$G.
}
\label{depsdE_and_s_vs_E}
\end{center}
\end{figure*}

In Sect.~\ref{sec:basics} we mentioned 
that the assumption of an energy-independent slope $s$ can introduce severe biases in the 
reliability of models and numerical simulations and ultimately affect the interpretation of synchrotron observations,
such as the spatial variation of $\beta$.
To prove this, we consider three observing frequencies (100~MHz, 1~GHz, and 10~GHz)
representative of all-sky Galactic radio surveys \citep{Platania+1998,Guzman+2011} 
and two extreme values for \bperp\ (2 and 20~$\mu$G) consistent with the magnetic field strength expected in the diffuse medium \citep{Heiles2005a,Beck2015,Ferriere2020}. 
To identify the range of energies contributing to the specific emissivity for a given value of $\nu$ and \bperp, we examine the integrands $d\varepsilon_{\nu}/d E$ summed over the two polarisations (Eqs.~(\ref{epsnu})). 
As shown in the upper panels of Fig.~\ref{depsdE_and_s_vs_E}, $d\varepsilon_{\nu}/d E$ 
has a well-defined maximum. For each frequency, we compute 
the energy range, around the energy of the maximum, where the
specific emissivity is equal to 95\% of its total.
This fiducial 95\% level is meant 
to show that most of the synchrotron emission at a given frequency originates from a definite, although broad, energy range
of CRe.
The resulting energy ranges (and the corresponding ranges of spectral slope $s$
derived for the CRe spectrum by \citealt{Orlando2018}) are listed in Table~\ref{tab:95}.
Looking at the ranges of $s$ at each frequency, it is clear that only in the extreme case of high frequencies 
($\nu\gtrsim10$~GHz) and very small \bperp\ ($\simeq2~\mu$G), the hypothesis of an energy-independent spectral slope 
is accurate (in this case $s\simeq-3.3$).
We note that the energy and the spectral slope intervals are the same for a given 
$\nu/\bperpmath$ ratio. This is because the total power per unit frequency, 
$P^{\rm em}_{\nu}$, emitted at a frequency $\nu$ by an 
electron of energy $E$ in a field \bperp, peaks at 
an energy proportional to $\sqrt{\nu/\bperpmath}$ \citep[see, e.g.,][]{Longair2011}.
Figure~\ref{depsdE_and_s_vs_E} shows that, for the typical frequency range of synchrotron observation in the diffuse ISM
and for the expected values of \bperp, 
CRe with energies between 100~MeV and 50~GeV are responsible for almost all the non-thermal emission.
Since in this energy range the spectral slope has the largest variation,
the assumption of an energy-independent $s$ is not correct.

For the sake of simplicity, we have only considered the case of the CRe spectrum by \citet{Orlando2018}. 
Using instead that of \citet{Padovani+2018a}, the energy intervals we obtain are similar, as the two spectra differ 
on average by 25\% (see Sect.~\ref{sec:basics} and Fig.~\ref{electronspectra}). 
However, what changes are the spectral slope intervals (see Fig.~\ref{energyslope}). 
This means that the use of a particular CRe spectrum has a major influence on the interpretation of 
the observations. In the next section we show how observational estimates of the
brightness temperature spectral index help to constrain the accuracy of a CRe spectrum.
\begin{table}[!h]
\caption{Ranges of energy $E$ and spectral slope $s$ 
for two values of \bperp\ and for a set of frequencies$^{a}$.}
\begin{center}
\resizebox{\linewidth}{!}{
\begin{tabular}{ccccc}
\toprule\toprule
$\nu$ [GHz]& \multicolumn{2}{c}{$\bperpmath=2~\mu$G} & \multicolumn{2}{c}{$\bperpmath=20~\mu$G}\\
\midrule
& $E$ [GeV] & $s$ & $E$ [GeV] & $s$\\
\cmidrule{2-5}
0.1 & [0.7,5.8] & [$-3.20,-1.94$] & [0.2,2.7] & [$-2.52,-1.86$]\\
1   & [2.1,14.6]& [$-3.27,-2.33$] & [0.7,5.8] & [$-3.20,-1.94$]\\
10  & [6.3,43.6]& [$-3.30,-3.24$] & [2.1,14.6]& [$-3.27,-2.33$]\\
\bottomrule
\end{tabular}
}
\end{center}
{\small
\begin{flushleft}
$^a$~~The ranges of $E$ and $s$ have been derived for the CRe spectrum by
\citet{Orlando2018} by integrating $d\varepsilon_{\nu}/d E$ in an 
energy range around the peak value until 95\% of the total specific 
emissivity is recovered.
\end{flushleft}
}
\label{tab:95}
\end{table}%

\section{Modelling synchrotron emission}
\label{sec:modelsynem}

In this section we first consider a cloud modelled as a simple uniform slab to show how specific realisations of the 
CRe spectrum can be ruled out by comparison with observational estimates of $\beta$ (Sect.~\ref{ssec:betaO18P18}).
We then describe the numerical simulations that we will use in the rest of the paper to study the spatial variations of $\beta$ and of the polarisation fraction (Sect.~\ref{ssec:sims}).

\subsection{Uniform slab}
\label{ssec:betaO18P18}

To show the importance of an accurate modelling of the 
CRe spectrum, we compute the brightness temperature, $T_\nu$,
for a slab with a 
fixed, spatially uniform, component of the magnetic field \bperp, varying from 2 to 20~$\mu$G
as 
specified in Sect.~\ref{sec:basics}, exposed to a flux of CRe, $j_e(E)$, given by the \citet{Orlando2018} and \citet{Padovani+2018a} spectra. 
For illustration, we select the frequency range 
$\nu=115$--189~MHz, with 
a frequency resolution of 183 kHz 
and assume an angular resolution of $\theta_b=4'$.
These parameters are representative
of the LOFAR High Band Antenna observations carried out by 
\citet{Jelic+2015}. We assume a thickness of the slab of 1~pc, 
although the brightness temperature can
be easily scaled for any different value.
We then compute the spectral index, $\beta$, through a linear fit of $\log T_{\nu}$
versus $\log\nu$ for each \bperp\ value. 

\begin{figure}[!h]
\begin{center}
\resizebox{0.9\hsize}{!}{\includegraphics{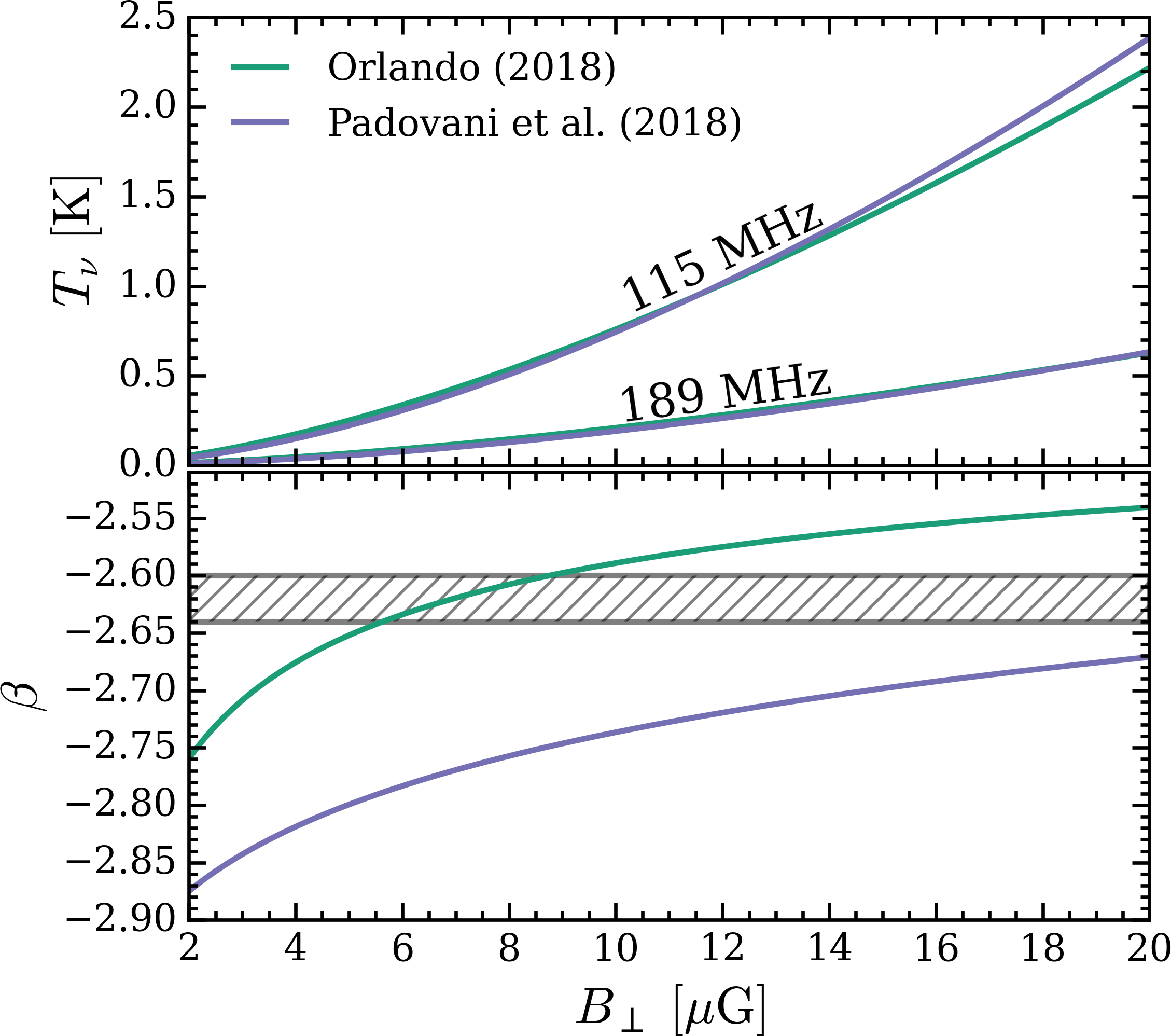}}
\caption{Upper panel: brightness temperature $T_\nu$ as a function of \bperp\, computed at $\nu=115$ and 189~MHz
with the CRe spectra by
\citet{Orlando2018} (green line) and \citet{Padovani+2018a} (violet line) for a slab of thickness 1~pc, with 
an angular resolution $\theta_b=4'$.
Lower panel: brightness temperature spectral index $\beta$ as a function of $B_\perp$
(same colour coding as in the upper panel).
The hatched region highlights typical values of 
$\beta$ at intermediate and high latitudes 
\citep{Mozdzen2017}.
}
\label{TbetaO18P18}
\end{center}
\end{figure}

The brightness 
temperatures generated by the two realisations of the CRe spectrum 
are on average within $\simeq 25$\% of each other  
(upper panel of Fig.~\ref{TbetaO18P18}), 
of the order of a few to several K. Assuming a synchrotron polarisation fraction of $p \approx 70$\% (more details in Sect.~\ref{ssec:polfrac}) the polarised intensity would be of the same order as $T_\nu$. This is exactly the amount of diffuse polarised emission 
observed by LOFAR in the range 100--200~MHz \citep[e.g.,][]{Jelic+2014,Jelic+2015,vanEck2017}. These simple arguments support the scenario, suggested by \citet{vanEck2017} and \citet{Bracco+2020}, where
pc-scale neutral clouds in the diffuse ISM 
could significantly contribute
to the Faraday-rotated synchrotron polarisation observed with LOFAR.
However, if we compare the value of $\beta$ 
(lower panel of Fig.~\ref{TbetaO18P18})
with typical values derived 
from observations at intermediate and high Galactic latitudes 
\citep[][shaded strip in the lower panel of Fig.~\ref{TbetaO18P18}]{Mozdzen2017}, it is evident that
in this example only the spectrum by \citet{Orlando2018} 
is consistent with the observations, and only for values 
of \bperp\ of the order of 6--9~$\mu$G, 
while the one by \citet{Padovani+2018a}
requires unrealistically high values of $B_\perp$.
As this example illustrates, the constraint imposed by
the highly accurate determination of the spectral 
index $\beta$ allowed by current instruments is a strong motivation to further and thoroughly model the synchrotron emission -- total and polarised -- of the diffuse ISM. 

\subsection{Numerical simulations}
\label{ssec:sims}

We now compute synthetic observations of synchrotron
emission adopting state-of-the-art numerical simulations of
the diffuse, 
magnetised, multiphase, turbulent, and neutral atomic ISM.
Our approach here is not to consider the simulations described below as true representations of any given LOS in the diffuse ISM, but rather as a laboratory to study realistic physical conditions of the multiphase medium where synchrotron emission may originate. In particular, we perform numerical simulations of the diffuse ISM using the RAMSES code \citep{Teyssier2002,Fromang2006}, a grid-based solver with adaptive mesh refinement
\citep{berger_adaptive_1984}, and a fully-treated tree data structure \citep{khokhlov_fully_1998}. 

The gas density of the medium we consider is typically dominated by neutral hydrogen
that can be traced via the 21 cm emission line \citep{heiles_millennium_2003,heiles_vizier_2003,murray_21-sponge_2015,murray_21-sponge_2018}. This line is usually decomposed into several Gaussian components  \citep{kalberla_properties_2018,marchal_rohsa_2019} associated to distinct gas phases in pressure balance: 
a dense cold neutral medium, CNM, with temperature and density  
$T\simeq50$~K and $n_{\rm H}\simeq50$~cm$^{-3}$, 
respectively, immersed in a diffuse warm neutral medium, WNM, with $T\simeq8000$~K and
$n_{\rm H}\simeq0.3$~cm$^{-3}$, 
and a third intermediate unstable phase, with temperature comprised between those of the CNM and the WNM \citep[e.g.][]{Wolfire2003,Bracco+2020}. 
\citet{field_thermal_1965} and
\citet{field_cosmic_1969} pointed out that the microphysical processes of heating and cooling naturally lead to two thermally stable (CNM and WNM) and a thermally unstable phases coexisting in a range of thermal pressure.
Through the thermal processes of condensation and evaporation, and with the help of turbulent transport and turbulent mixing, the diffuse matter can flow from one stable state to the other. 

The local diffuse matter in our Galaxy is simulated over a box of 50~pc with periodic boundary conditions,
using a fixed grid of $128^{3}$ pixels, corresponding to
an effective resolution of 0.39~pc. 
The initial state is characterised by
a homogeneous density  
$n_{\rm H}=$1.5~cm$^{-3}$, a temperature $T=8000$~K, and a uniform magnetic field
$\vec{B}_{0}=B_{0}\hat{e}_{x}$.
We consider two snapshots of the simulation, 
one with a standard average magnetic field strength, $B_{0}=4~\mu$G
(hereafter ``weak field'' case),
and one with a stronger $B_{0}=20~\mu$G
(hereafter ``strong field'' case).
The gas evolves under the joint influence of turbulence, magnetic field, radiation field, and thermal instability, 
and separates in three different phases: CNM, WNM, and unstable.
A turbulent forcing is applied to mimic the injection of mechanical energy in the diffuse ISM. 
Following \citet{schmidt_numerical_2009} and \citet{federrath_comparing_2010}, this forcing, modelled by an acceleration term in the momentum equation, is driven through a pseudo-spectral method. 
The turbulent acceleration parameter is set to $2.77\times10^{-3}$~kpc~Myr$^{-2}$.
Using classical notation, the dimensionless compressible parameter for the turbulent forcing modes, 
$\zeta_{0}$, ranges from pure solenoidal modes ($\zeta_{0}=1$) to pure compressible modes ($\zeta_{0}=0$).
We set $\zeta_{0}=0.5$.

The matter is assumed to be illuminated on all sides by an isotropic spectrum of UV photons set to the standard interstellar radiation field \citep{Habing1968}.
To correctly describe the thermal state of the diffuse ISM, we have included
the heating induced 
by the photoelectric effect on interstellar dust grains and secondary electrons produced during 
cosmic-ray propagation, and the cooling induced by emission of Lyman-$\alpha$ photons, the fine-structure lines of OI and CII, and the recombination of electrons onto grains. All these processes, described in Appendix B of \citet{Bellomi+2020}, are modelled with the analytical formulae given by \citet{Wolfire2003}.

These simulations represent two different scenarios:
in the ``weak field'' case, the magnetic field 
has a turbulent 
component of the same order of its mean component.
By contrast, in the ``strong field'' case, the 
magnetic field is mostly directed along the $x$-axis, i.e. its projection is mainly 
contained in the $xy$ and $zx$ planes.
The advantage of using simulations is that we can rotate each snapshot according to the three axes and 
calculate the quantities of interest 
integrated along three different LOS, 
thus increasing our statistics.
Table~\ref{tab:s} summarises 
the ranges of the strength of the magnetic field in the plane perpendicular to a 
given LOS
and their median values (marked by a 
superscript tilde) for the two snapshots under
consideration. 
Specifically, for the LOS $i$, we compute the minimum and maximum value of the magnetic field strength, $B_{jk}=(B_{j}^{2}+B_{k}^{2})^{1/2}$,
in the plane of the sky (POS) $jk$.
Subscripts $i$, $j$, and $k$ follow the cyclic permutation of Cartesian coordinates 
$(ijk)=\{xyz,yzx,zxy\}$.
\begin{table}[!h]
\caption{Minimum and maximum values of the strength of the magnetic field in the plane perpendicular
to the LOS $i$, $B_{jk}=(B_{j}^{2}+B_{k}^{2})^{1/2}$, and its median value
 $\tilde B_{jk}$  for
the two cases of weak and strong field$^{a}$. 
}
\begin{center}
\begin{tabular}{ccccc}
\toprule\toprule
&LOS$_{i}$ & $B_{jk}$~[$\mu$G] & $\tilde B_{jk}$~[$\mu$G]\\
\midrule
\multirow{3}{*}{\rotatebox[origin=c]{0}{\parbox[c]{2cm}{\centering weak field ($B_{0}=4~\mu$G)}}}
&$x$ & [$1.1\times10^{-3},12$]&$2.02^{+0.88}_{-0.75}$\\[3px]
&$y$ & [$1.5\times10^{-3},16$]&$4.26^{+0.73}_{-0.71}$\\[3px]
&$z$ & [$9.8\times10^{-3},13$]&$4.22^{+0.71}_{-0.74}$\\[3px]
\midrule
\multirow{3}{*}{\rotatebox[origin=c]{0}{\parbox[c]{2cm}{\centering strong field ($B_{0}=20~\mu$G)}}}
&$x$ & [$3.6\times10^{-4},6$]&$0.80^{+0.38}_{-0.30}$\\[3px]
&$y$ & [$17,22$]&$19.19^{+0.33}_{-0.41}$\\[3px]
&$z$ & [$17,22$]&$19.19^{+0.33}_{-0.40}$\\[3px]
\bottomrule
\end{tabular}
\end{center}
{\small
\begin{flushleft}
$^a$~~Errors on $\tilde B_{jk}$ are estimated using the first and third quartiles.
Subscripts $i$, $j$, and $k$ follow the cyclic permutation of Cartesian coordinates
$(ijk)=\{xyz,yzx,zxy\}$.
\end{flushleft}
}
\label{tab:s}
\end{table}%

\section{Results}
\label{sec:results}

In contrast to typical angular
resolutions of earlier facilities, such as the $5^\circ$ of \citet{Guzman+2011}, nowadays, LOFAR provides a significantly higher resolution, typically up to $4'$ at frequencies of 115$-$189~MHz 
for observations of Galactic diffuse emission \citep{Jelic+2014,Jelic+2015,vanEck2017}.
In the following, we show our results 
at the resolution of 
$6.7'$. 
The latter value 
corresponds to the spatial resolution of the simulations (0.39~pc)
if the simulation snapshots are placed at a distance of 200~pc.
We focus on the frequency range 115$-$189~MHz, 
however the same conclusions apply to higher frequencies 
(see Sect.~\ref{ssec:lookup} and Appendix~\ref{app:betaLMH}).

\subsection{Brightness temperature maps}
\label{ssec:Tmaps}

\begin{figure}[!t]
\begin{center}
\resizebox{1\hsize}{!}{\includegraphics{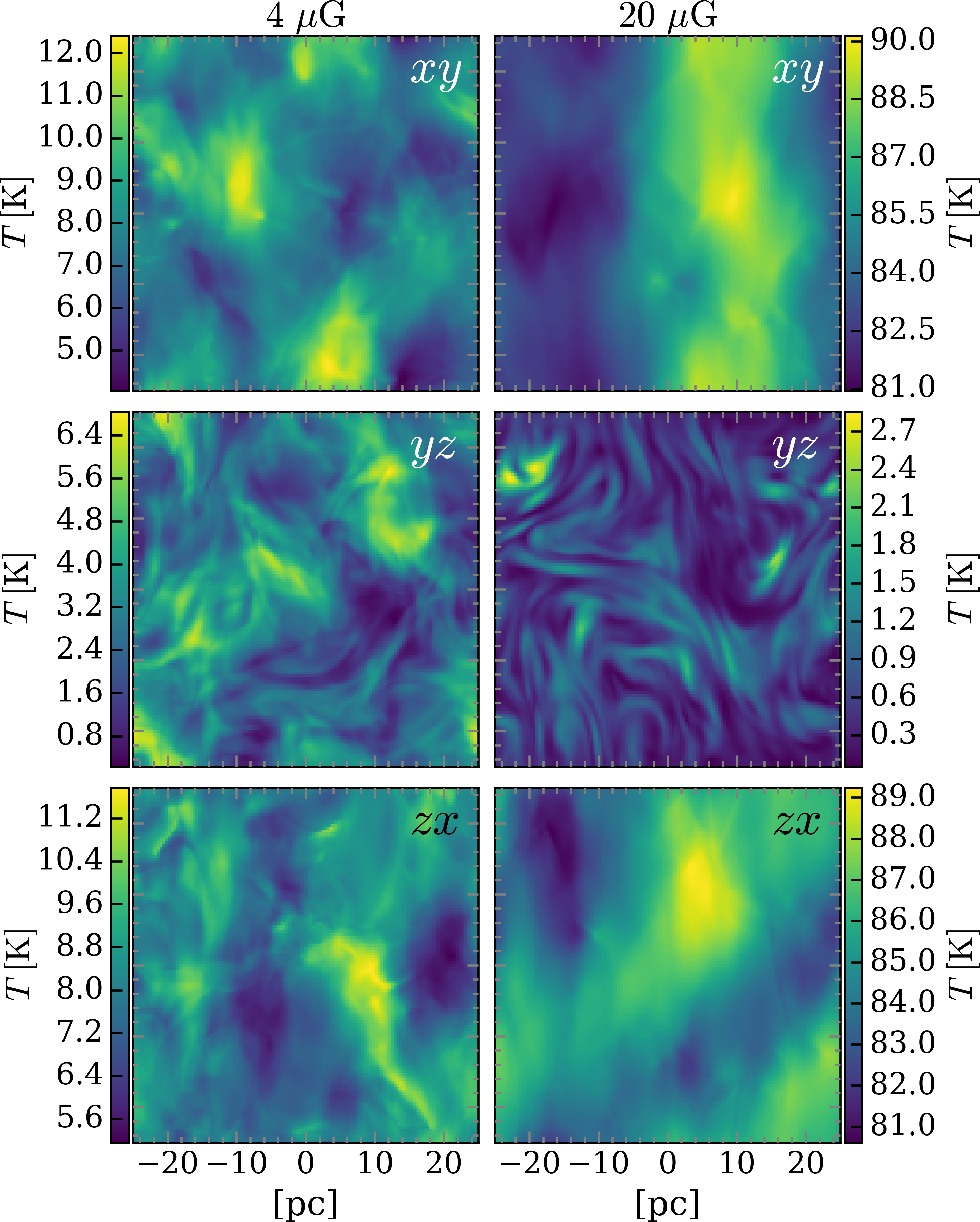}}
\caption{Brightness
temperature maps (colour scale) 
computed with the \citet{Orlando2018} spectrum, 
at $\nu=130$~MHz and with resolution $6.7'$ for the three POS (three rows,
labelled $xy$, $yz$, and $zx$) of the weak and strong field case (left and right column,
respectively).}
\label{T130MHz}
\end{center}
\end{figure}

Figure~\ref{T130MHz} shows the brightness temperature maps
at a frequency $\nu=130$~MHz 
for the two snapshots
described in Sect.~\ref{ssec:sims}, obtained with  
the CRe spectrum of \citet{Orlando2018}.%
\footnote{We verified that both 
the effect of synchrotron self-absorption
\citep{GinzburgSyrovatskii1965} 
and the Tsytovich-Razin effect
\citep{GinzburgSyrovatskii1964}
can be neglected for the ISM under
consideration.}
These images are derived by integrating along the
three different LOS ($x$, $y$, $z$) over the length of the snapshot (50~pc), 
resulting in temperature maps identified by the respective POS ($yz$, $zx$, $xy$).
As anticipated by the upper panel of Fig.~\ref{TbetaO18P18},
the temperature is higher where \bperp\ is larger, 
reaching a maximum in the $xy$ and $zx$ POS of the strong field case (right column of Fig.~\ref{T130MHz}; 
see also the values of $\tilde B_{jk}$ in Table~\ref{tab:s}), while
the smallest temperature values are in the $yz$ POS of the strong field case.
Temperature maps are more inhomogeneous in the $yz$ POS of both snapshots. 
That is because in these planes \bperp\ has a significant turbulent component. 
Therefore, we introduce 
the ratio between the standard deviation and the median value of
\bperp, $\mu=\sigma_{\bperpmath}/\tilde{B}_{\perp}$,
to quantify the relative turbulent component of \bperp.
This quantity, whose median values are reported in Table~\ref{tab:betap},
is useful for the discussion on $\beta$ and on the polarisation fraction
in the following two subsections.

\subsection{Brightness temperature spectral index}
\label{ssec:beta}

\begin{figure}[!h]
\begin{center}
\resizebox{1\hsize}{!}{\includegraphics{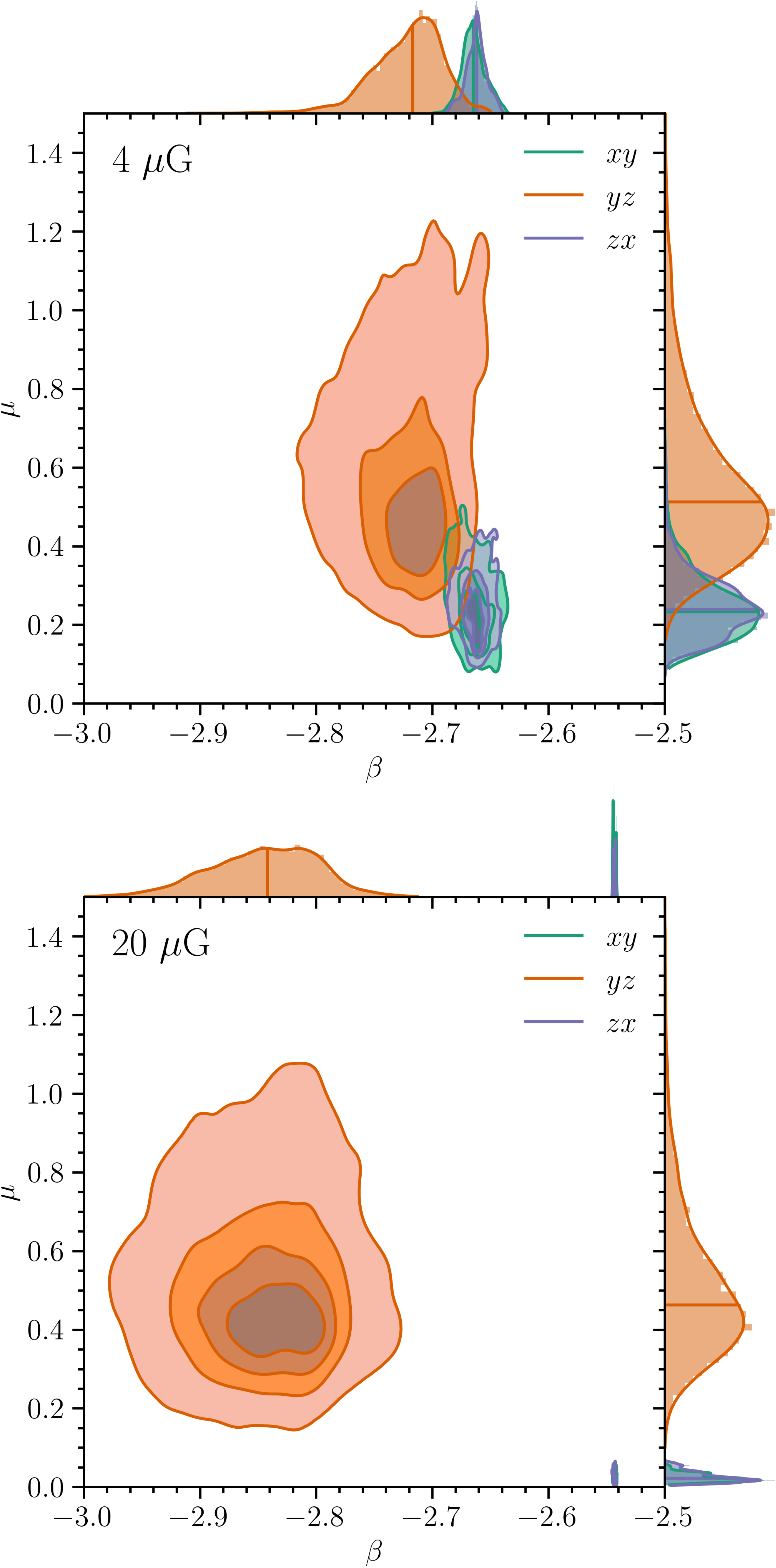}}
\caption{
Bivariate distribution of the ratio between
the standard deviation of $B_\perp$ and its median value, $\mu$, and
the brightness temperature spectral index, $\beta$, computed 
in the frequency range 115$-$189~MHz
for each LOS
for the weak and strong field case (upper and lower panel, respectively),
at a resolution of $6.7'$.
The POS are identified by the 
three different colours displayed in the legend.
Isodensity contours are plotted at 5\%, 30\%, 50\%, and 75\% levels.
The histograms above and to the right of the main plots show the 
marginal distribution of the two quantities. 
The median values are marked by horizontal and vertical lines 
and listed in Table~\ref{tab:betap}. }
\label{histobeta}
\end{center}
\end{figure}

We compute the brightness temperature spectral index, $\beta$, 
for each LOS, by a linear fit of $\log T_{\nu}$ versus $\log\nu$
for the three POS of the two snapshots, 
in the frequency range 
115--189~MHz with 
a frequency resolution of 183 kHz
as in the LOFAR dataset by
\citet{Jelic+2015}.
In Fig.~\ref{histobeta}, we show the results
in the form of bivariate and marginal distributions 
(see also Appendix~\ref{app:beta} for the maps of $\beta$). 
From the inspection of Fig.~\ref{histobeta}, it is evident that in no case, 
at low frequencies, is $\beta$ equal to $-2.5$, as would follow from the assumption of constant
$s=-2$. As shown in Sect.~\ref{ssec:srange}, the same consideration holds in the high frequency regime 
($\nu\gtrsim400$~MHz), where the assumption of constant $s=-3$, and hence $\beta=-3$, 
turns out to be incorrect.
Figure~\ref{histobeta} also shows that $\tilde\beta$ is less negative in the POS $xy$ and $zx$, 
where $\tilde B_\perp$ is larger
(see also Tables~\ref{tab:s} and~\ref{tab:betap}). 
This is a consequence of what is shown in Fig.~\ref{depsdE_and_s_vs_E}: 
for a given frequency, the larger the value of \bperp\ at a given position along the LOS, the smaller the median value of the energy range determining its synchrotron emissivity. 
Therefore, the corresponding values of $s$, and hence of $\beta$, increase.
Finally, an important aspect to note is that the dispersion of $\beta$ values around its median 
depends weakly on $\mu$. Indeed, in both $yz$ POS, where $\mu$ reaches the largest values, 
the interquartile range (i.e. the difference between the third and the first quartile) is 
only 0.04 and 0.06 
for the weak and the strong case, respectively. 
For the sake of completeness, in Appendix~\ref{app:betaLMH} we show the bivariate distributions of 
$\mu$ and $\beta$ for higher frequency ranges (467$-$672~MHz and 833$-$1200~MHz), 
which will be used later in Sect.~\ref{ssec:lookup}. 
The conclusions are the same as for the 115$-$189~MHz range.

\begin{table}[!h]
    \caption{ 
    Medians of 
    the ratio between the standard deviation of $B_\perp$ and its
    median value, $\mu=\sigma_{B_\perp}/\tilde B_\perp$,
    of the brightness temperature spectral index, 
    and of the polarisation fraction at a resolution of $6.7'$
for the three POS and the two cases of weak and strong field$^{a}$.} 

\begin{center}
\resizebox{\linewidth}{!}{
\begin{tabular}{ccccc}
\toprule\toprule
&POS & $\tilde\mu$ & $\tilde\beta$ & $\tilde p$\\
\midrule
\multirow{3}{*}{\rotatebox[origin=c]{0}{\parbox[c]{2cm}{\centering weak field ($B_{0}=4~\mu$G)}}}
&$xy$ & $0.23^{+0.06}_{-0.05}$ & $-2.66\pm0.01$ & $0.57^{+0.05}_{-0.06}$\\[3px] 
&$yz$ & $0.51^{+0.14}_{-0.10}$ & $-2.72\pm0.02$ & $0.34\pm0.10$\\[3px]
&$zx$ & $0.24^{+0.05}_{-0.05}$ & $-2.66\pm0.01$ & $0.55\pm0.06$\\
\midrule
\multirow{3}{*}{\rotatebox[origin=c]{0}{\parbox[c]{2cm}{\centering strong field ($B_{0}=20~\mu$G)}}}
&$xy$ & $0.02\pm0.01$ & $-2.54$ & $0.70$\\
&$yz$ & $0.47^{+0.12}_{-0.09}$ & $-2.84\pm0.03$ & $0.51^{+0.09}_{-0.13}$\\[3px]
&$zx$ & $0.02\pm0.01$ & $-2.54$ & $0.70$\\
\bottomrule
\end{tabular}
}
\end{center}
{\small
\begin{flushleft}
$^a$~~Errors have
been estimated using the first and third quartiles.
Errors smaller than 0.01 are not shown.
\end{flushleft}
}    
\label{tab:betap}
\end{table}



\subsection{Polarisation fraction}
\label{ssec:polfrac}

\begin{figure}[!h]
\begin{center}
\resizebox{1\hsize}{!}{\includegraphics{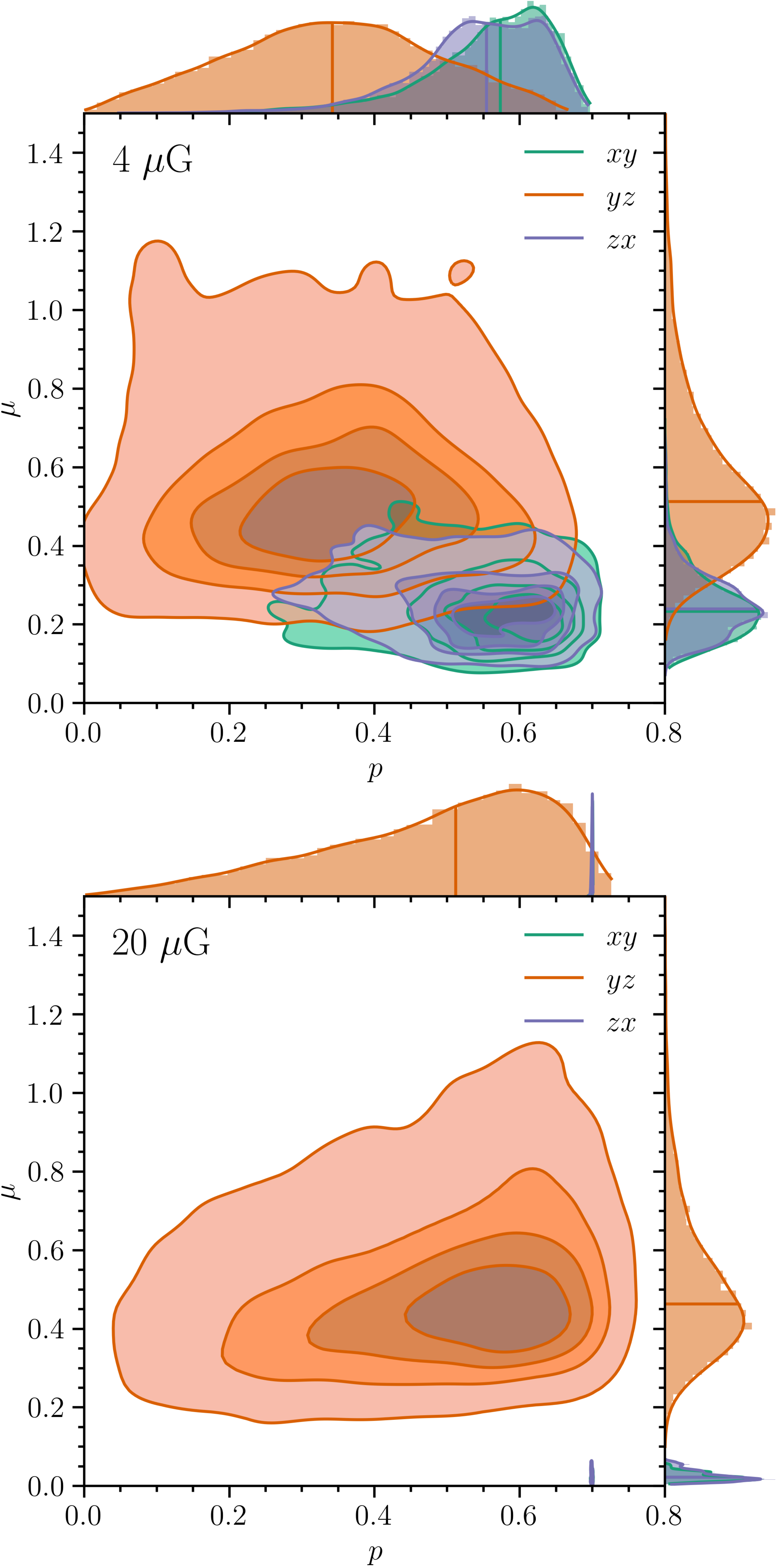}}
\caption{
Bivariate distribution of the ratio between
the standard deviation of $B_\perp$ and its median value, $\mu$, and
the polarisation fraction, $p$, computed at 130~MHz for each LOS
for the weak and strong field case (upper and lower panel, respectively),
at a resolution of $6.7'$.
The POS are identified by the 
three different colours displayed in the legend.
Isodensity contours are plotted at 5\%, 30\%, 50\%, and 75\% levels.
The histograms above and to the right of the main plots show the 
marginal distribution of the two quantities. 
The median values are marked by horizontal and vertical lines 
and listed in Table~\ref{tab:betap}.}
\label{histop}
\end{center}
\end{figure}

The local polarisation fraction, i.e., the polarisation fraction based
on the local emissivities for an energy-independent value of $s$ is
\be\label{defp}
p=\frac{3-3s}{7-3s}=\frac{3+3\beta}{1+3\beta}\,,
\ee
\citep[see, e.g.][]{RybickiLightman86}, which, for $s=-2$ and $-3$, gives $p=69\%$ 
and 75\%, respectively.
If the orientation and the strength of \vecbperp\ 
do not vary along the LOS, 
Eq.~(\ref{defp}) would also give the polarisation fraction in the POS.
To quantify the limitations of the above assumption, 
we calculate the expected polarisation fraction (Eq.~(\ref{polfraceq}))
at the reference frequency $\nu=130$~MHz for the two simulations and at the resolution of $6.7'$
in the three POS.
Similarly to Fig.~\ref{histobeta}, in Fig.~\ref{histop}
we show the results in the form of bivariate and marginal distributions, and we find a similar correlation of
$\mu$ with $p$ as we found 
with $\beta$: as $\mu$ increases,
the spread of values of $p$ around its median value 
is even larger than that of $\beta$ (see Table~\ref{tab:betap}). 

Furthermore, except for
the POS $xy$ and $zx$ in the strong field case,
the median value is much lower than the  
theoretical value of $\simeq70\%$ of the local polarisation fraction, which is the 
maximum value of the observed polarisation fraction. 
Such depolarisation effect in our study is mainly caused by the tangling of the turbulent component of the magnetic field along the LOS. This effect has been extensively reported in the literature both in the radio \citep[e.g.,][]{Gaensler2011} and in the sub-millimetre domain \citep[e.g.,][]{PlanckXX2015}. Moreover, radio synchrotron polarisation can be severely affected by  other mechanisms that drastically reduce the amount of detectable polarised emission, such as Faraday rotation and beam depolarisation \citep[see, e.g.,][]{Sokoloff1998,Haverkorn2004}. We note that Faraday rotation at 
a few hundred MHz cannot be neglected from an observational point of view. However, in this study we choose not to detail this process and to only focus on the emission mechanism of synchrotron radiation with an 
energy-dependent spectral slope. Faraday rotation will be the core of a follow-up paper that will dig into the complexity of the ionisation degree of the multiphase diffuse ISM, a key ingredient to correctly model the effect of Faraday rotation.

\section{Discussion}
\label{sec:discussion}

\subsection{Effect of angular resolution on $\beta$ and $p$}
\label{ssec:resolution}
\begin{figure}[!h]
\begin{center}
\resizebox{1\hsize}{!}{\includegraphics{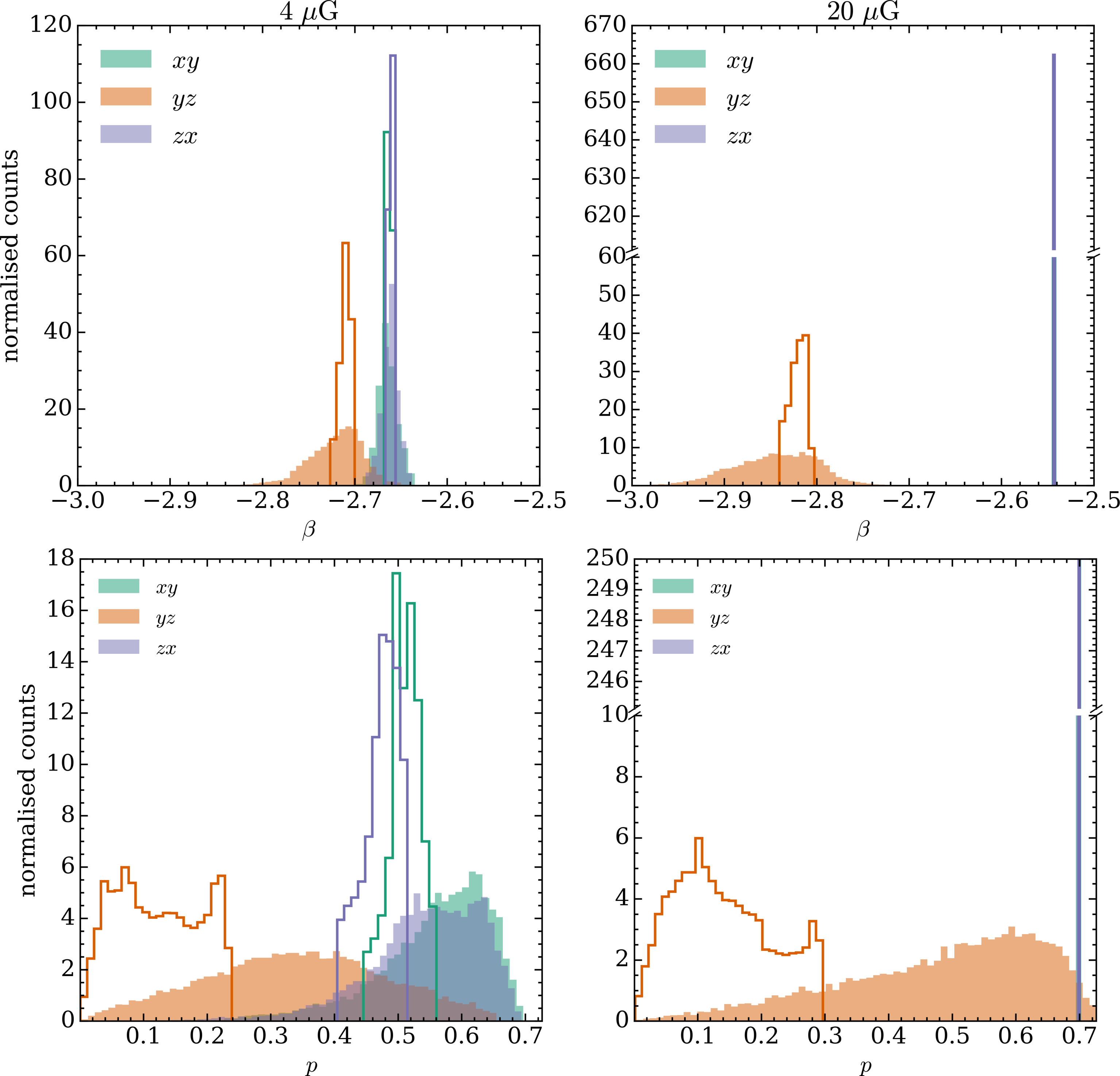}}
\caption{Histograms of the spectral index 
$\beta$ (upper panels) and
polarisation fraction $p$ (lower panels)
for the weak and strong field case (left and right column,
respectively). Solid (empty) histograms refer to a
resolution of $6.7'$ ($5^\circ$). The POS are
identified by the three different colours displayed in the 
legend.}
\label{histo2res}
\end{center}
\end{figure}

In this study we have assumed the typical angular resolution of observations 
carried out with the most recent facilities, such as LOFAR. 
Here we show the distributions of
the spectral index and the polarisation fraction obtained from lower resolution observations such as those shown in 
\citet{Guzman+2011}. The latter paper 
presented an all-sky Galactic radio emission map,
convolved to a common $5^\circ$ resolution,
combining observations obtained with different telescopes such as the Parkes 64m, the Jodrell Bank MkI 
and MkIA, and the Effelsberg 100m \citep{Haslam+1981,Haslam+1982}, the Maip\'u Radio Astronomy 
Observatory 45-MHz array \citep{Alvarez+1997}, and the Japanese Middle and Upper atmosphere radar array \citep[MU radar;][]{Maeda+1999}. 
Figure~\ref{histo2res} shows the comparison of the distributions of
$\beta$ and $p$ 
(upper and lower panels, respectively)
at the resolutions of $6.7'$ and $5^\circ$
(filled and empty histograms, respectively). 
Despite the increase in resolution of LOFAR, the median value of $\beta$ 
does not change appreciably (see upper panels). However, what does change is 
the spread of the distribution, since at a lower resolution it is not possible 
to pick up the finer variations of the spectral index. 
More interesting is how the polarisation fraction distributions differ 
in the two cases. 
The lower panels show the beam depolarisation effect
mentioned in Sect.~\ref{ssec:polfrac}:
where $\mu$ 
is larger, a lower resolution clearly results in a lower $p$.

\subsection{A look-up plot for $B_\perp$}
\label{ssec:lookup}

At any given position along a LOS, the synchrotron emission is jointly determined by the local CRe spectrum 
and the local value of \bperp.
As we showed in Sect.~\ref{ssec:betaO18P18}, thanks to the range of $\beta$ obtained from observations, 
it is possible to constrain the 
CRe spectrum at $\sim$GeV energies 
(see also Fig.~\ref{depsdE_and_s_vs_E}). This 
makes it possible
to reduce the uncertainly on the spectral shape, 
an appreciable improvement over the assumption of an 
energy-independent $s$.
One should also 
consider that the latest generation of telescopes, such as LOFAR and in the near future SKA, reach resolutions up to at least
three orders of magnitude higher than previous instruments.
This means that the uncertainties on the values of $\beta$ derived from observations
will be reduced and it will be possible to obtain $\beta$ variations over smaller fields of view.
Finally, in Sect.~\ref{ssec:beta} 
we showed that variations of \bperp\ along the LOS do not affect the estimates of $\beta$.
This feature is particularly relevant for reducing 
the uncertainty in the estimation of the average value of \bperp\ along a LOS.
Given these premises, we suggest a method for estimating $\langle\bperpmath\rangle$, the average value of \bperp,
along a given LOS.

\begin{figure}[!h]
\begin{center}
\resizebox{1\hsize}{!}{\includegraphics{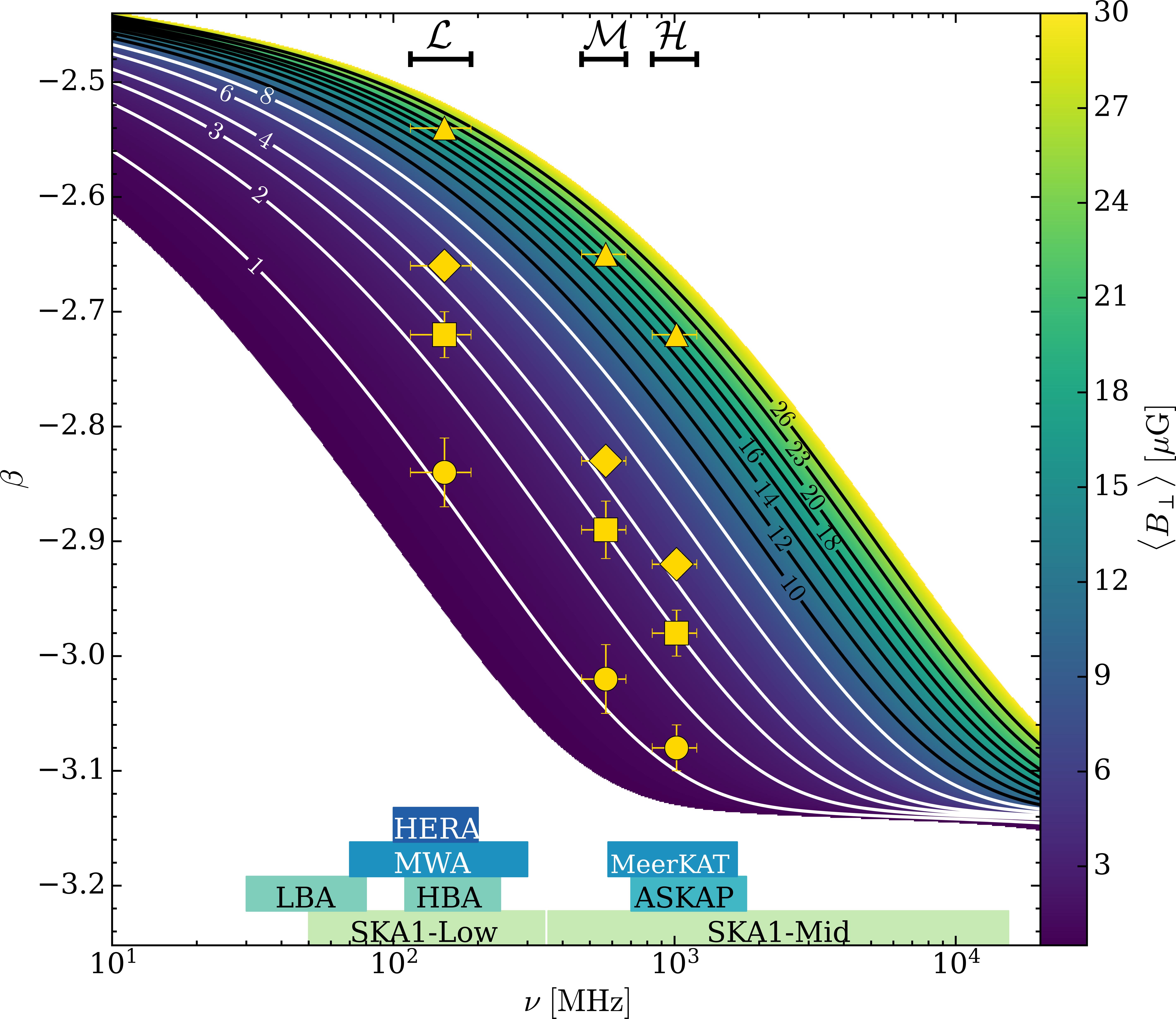}}
\caption{A look-up plot to estimate 
the average value of \bperp\ along a LOS, $\langle\bperpmath\rangle$,
for a given frequency $\nu$ and 
brightness
temperature spectral index $\beta$
assuming the CRe spectrum by \citet{Orlando2018}. Black and white isocontours show
specific values of $\langle\bperpmath\rangle$ with labels in $\mu$G.
Yellow symbols indicate estimates of $\langle\bperpmath\rangle$ obtained from
the values of $\tilde\beta$ in the frequency ranges 
115$-$189~MHz, 467$-$672~MHz, and 833$-$1200~MHz  
(labelled as $\mathcal{L}$, $\mathcal{M}$, and $\mathcal{H}$, respectively),
at a common resolution of $6.7'$,
for the
different POS of the two simulation snapshots 
(circles: POS $yz$ in the strong field case;
squares: POS $yz$ in the weak field case;
diamonds: POS $xy$ and $zx$ in the weak field case;
triangles: POS $xy$ and $zx$ in the strong field case).
Rectangles at the bottom of 
the plot show the frequency ranges of 
the low and high band antenna of LOFAR (LBA and HBA, respectively),
the low and mid frequency bands
of SKA (SKA1-Low and SKA1-Mid, respectively), and 
SKA pathfinders and precursors:
the Hydrogen Epoch or Reionisation (HERA);
the Murchison Widefield Array (MWA);
the Karoo Array Telescope (MeerKAT);
the Australian SKA Pathfinder (ASKAP).}
\label{lookupplot}
\end{center}
\end{figure}

We adopt the \citet{Orlando2018} CRe spectrum and assume $\nu$ in the range 10~MHz$-$20~GHz
and $\langle\bperpmath\rangle$ in the range 0.5$-$30~$\mu$G. 
We compute the corresponding emissivities 
(see Eqs.~(\ref{epsnu})), integrated along the LOS for 1~pc,%
\footnote{Since we are interested in $\beta$, results are independent of the LOS path length.} 
and compute $\beta(\nu,\langle\bperpmath\rangle)=d\log T_{\nu}(\langle\bperpmath\rangle)/d\log\nu$.
By inverting the relation for $\beta$, 
in Fig.~\ref{lookupplot} we show the values of $\langle\bperpmath\rangle$ 
expected for each couple $(\nu,\beta)$.

We then consider three different ranges of frequencies:
115$-$189~MHz with a resolution of 183~kHz, 
467$-$672~MHz with a resolution of 507~kHz,
and 833$-$1200~MHz with a resolution of 908~kHz,%
\footnote{Spectral resolutions were chosen so as to have the same number of frequency bins for the three frequency intervals considered.}
labelled in Fig.~\ref{lookupplot}
as low ($\mathcal{L}$), mid ($\mathcal{M}$), and high ($\mathcal{H}$) frequency range, respectively.
Following the procedure described in Sect.~\ref{ssec:beta}, 
for each POS of the two snapshots, we calculate the temperature maps as a function of frequency
at a resolution of $6.7'$  
and we extract the $\beta$ value for each LOS and for each of the three frequency intervals,
$\mathcal{L}$, $\mathcal{M}$, and $\mathcal{H}$,
by a linear fit of $\log T_{\nu}$
versus $\log\nu$.
We show in Fig.~\ref{lookupplot} the $\tilde\beta$ values for each POS by estimating the errors using 
the first and third quartiles for each frequency interval. We note that, since the POS $xy$ and $zx$ 
for both snapshots have the same $\tilde\beta$, in the figure we show four values instead of six.
The values of $\tilde\beta$ for the $\mathcal{L}$ frequency range are listed in Table~\ref{tab:betap},
while those for the $\mathcal{M}$ and $\mathcal{H}$ ranges are reported in Table~\ref{tab:betaLMH}.

Figure~\ref{lookupplot} shows that, for each POS, the estimates of $\tilde\beta$ in the three frequency 
intervals correspond to the same $\langle\bperpmath\rangle$, which also agree
with the respective value of $\tilde{B}_{jk}$ listed in Table~\ref{tab:95}.
This confirms the consistency of our procedure.

This plot shows that there is a preferred range of frequencies that can be conveniently used to estimate 
$\langle\bperpmath\rangle$ along a LOS. 
This range corresponds approximately to 100~MHz$-$5~GHz, 
the frequency interval where $s$, and hence $\beta$, 
varies the most and where the isocontours of $\langle\bperpmath\rangle$ 
are more separated. 
Conversely, at frequencies that are too low (high), the CRe spectrum flattens out if 
$\langle\bperpmath\rangle$ is too high (low), and the estimate of $\langle\bperpmath\rangle$
becomes more uncertain.
From this figure it can be concluded that, in order to have a more precise estimate of 
$\langle\bperpmath\rangle$, it is advisable to  
simultaneously observe in narrow frequency ranges with high spectral resolution
(as our $\mathcal{L}$, $\mathcal{M}$, and $\mathcal{H}$ intervals)
in order to have independent $\beta$ estimates that 
should follow a specific isocontour of \bperp.

\section{Summary}
\label{sec:summary}

We carried out a quantitative study to understand the
consequences of an energy-dependent CRe spectral slope in the 
interpretation of observations of synchrotron emission in the diffuse and 
magnetised ISM. We focused in particular on metre 
wavelengths that can currently be observed with state-of-the-art facilities such as LOFAR
and in the near future with SKA.
At frequencies lower (higher) than
$\simeq 400$~MHz a constant spectral slope $s=-2$ ($s=-3$) is 
often assumed, mainly to avoid time-consuming calculations in analytical 
models and numerical simulations. 
As a consequence, 
one should also expect a constant value for the 
brightness temperature spectral index, $\beta$, related to $s$ by 
$\beta=(s-3)/2$ \citep{RybickiLightman86}.
However, metre wavelength observations show 
that $\beta$ is not constant across the Galaxy, taking on values quite 
different from $-2.5$ (corresponding to $s=-2$), varying between 
about $-2.7$ and $-2.1$ \citep{Guzman+2011}. 

For typical magnetic field strengths 
expected in the diffuse ISM ($\simeq2-20~\mu$G), 
the electrons that mostly determine the synchrotron emission 
at frequencies between about 100~MHz and 10 GHz
have energies 
in the range $\simeq$100~MeV$-$50~GeV. It is precisely at these energies that the spectral slope shows 
the largest variations. For example, for the  CRe spectrum described in \citet{Orlando2018},
representative of intermediate Galactic latitudes including
most of the local emission within about 1~kpc around the Sun,
in this energy range $s$ varies between about $-3.2$ and $-1.8$.

In order to understand the effect of an energy-dependent 
spectral slope at a
quantitative level, 
we first considered a slab with a fixed, spatially uniform \bperp\
exposed to a flux of CRe and we showed that, thanks to high-precision observational 
estimates of $\beta$ at low frequencies \citep{Mozdzen2017}, 
it is possible to discard some realisations of the CRe spectrum that 
would imply unrealistically high values of \bperp.
Then, we used two snapshots of 3D MHD simulations 
\citep{Bellomi+2020} with
different median magnetic field strength,
and studied the synchrotron emission according to the CRe spectrum by \citet{Orlando2018}.
We computed the distribution of $\beta$ for three frequency ranges, at low, mid, and
high frequencies (115$-$189~MHz, 467$-$672~MHz, and 833$-$1200~MHz, respectively).
We showed that the assumption of an energy-independent $s$ is 
not justified and leads to non-negligible biases in the interpretation of the observed 
spectral index distributions.
In particular, we found that $\beta$ becomes 
less negative as \bperp\ increases 
and that the
dispersion of the distribution of spectral index values around its median
weakly depends on how much \bperp\ varies along the LOS.
This property is of special relevance since, once a CRe spectrum is assumed, the uncertainty about the 
expected average value of \bperp
for a given LOS, $\langle\bperpmath\rangle$, is reduced.
We then presented a look-up plot that makes it possible to 
estimate $\langle\bperpmath\rangle$
given $\beta$ values obtained from observations in one or more frequency intervals. 
More precisely, we suggest repeating observations in narrow frequency intervals 
with high spectral resolution in order to have independent estimates of $\beta$ 
that should lie on the same $\langle\bperpmath\rangle$ isocontour in the look-up plot.

Finally, 
we computed the expected polarisation fraction, $p$, finding that
it is expected to be 
smaller the more turbulent
the magnetic field 
along the LOS is, deviating noticeably from 
the maximum value of $\simeq70\%$. 
The dispersion of $p$ around its median value is larger 
than that of $\beta$ as a consequence of the
turbulent tangling of the magnetic field lines along the LOS. 
The analysis of this depolarisation
effect and the consequences of Faraday rotation are 
deferred to a subsequent paper.

\begin{acknowledgements}
The authors wish to thank the referee, Katia Ferri\`ere, for her careful reading of the manuscript and insightful comments that considerably helped to improve the paper.
MP thanks Tommaso Grassi for useful discussions.
AB acknowledges the support from the European Union’s Horizon 2020 research and innovation program under the Marie Skłodowska-Curie Grant agreement No. 843008 (MUSICA). VJ acknowledges support by the Croatian Science Foundation for the project IP-2018-01-2889 (LowFreqCRO).
\end{acknowledgements}

\bibliographystyle{aa} 
\bibliography{mybibliography-bibdesk.bib} 

\begin{thebibliography}{66}
\expandafter\ifx\csname natexlab\endcsname\relax\def\natexlab#1{#1}\fi

\bibitem[{{Ackermann} {et~al.}(2010){Ackermann}, {Ajello}, {Atwood}, {Baldini},
  {Ballet}, {Barbiellini}, {Bastieri}, {Baughman}, {Bechtol}, {Bellardi},
  {Bellazzini}, {Belli}, {Berenji}, {Blandford}, {Bloom}, {Bogart},
  {Bonamente}, {Borgland}, {Brandt}, {Bregeon}, {Brez}, {Brigida}, {Bruel},
  {Buehler}, {Burnett}, {Busetto}, {Buson}, {Caliandro}, {Cameron}, {Caraveo},
  {Carlson}, {Carrigan}, {Casandjian}, {Ceccanti}, {Cecchi}, {{\c{C}}elik},
  {Charles}, {Chekhtman}, {Cheung}, {Chiang}, {Cillis}, {Ciprini}, {Claus},
  {Cohen-Tanugi}, {Conrad}, {Corbet}, {Deklotz}, {Dermer}, {de Angelis}, {de
  Palma}, {Digel}, {di Bernardo}, {Do Couto E Silva}, {Drell}, {Drlica-Wagner},
  {Dubois}, {Fabiani}, {Favuzzi}, {Fegan}, {Fortin}, {Fukazawa}, {Funk},
  {Fusco}, {Gaggero}, {Gargano}, {Gasparrini}, {Gehrels}, {Germani},
  {Giglietto}, {Giommi}, {Giordano}, {Giroletti}, {Glanzman}, {Godfrey},
  {Grasso}, {Grenier}, {Grondin}, {Grove}, {Guiriec}, {Gustafsson}, {Hadasch},
  {Harding}, {Hayashida}, {Hays}, {Horan}, {Hughes}, {J{\'o}hannesson},
  {Johnson}, {Johnson}, {Johnson}, {Kamae}, {Katagiri}, {Kataoka}, {Kerr},
  {Kn{\"o}dlseder}, {Kuss}, {Lande}, {Latronico}, {Lemoine-Goumard}, {Llena
  Garde}, {Longo}, {Loparco}, {Lott}, {Lovellette}, {Lubrano}, {Makeev},
  {Mazziotta}, {McEnery}, {Mehault}, {Michelson}, {Minuti}, {Mitthumsiri},
  {Mizuno}, {Moiseev}, {Monte}, {Monzani}, {Moretti}, {Morselli}, {Moskalenko},
  {Murgia}, {Nakamori}, {Naumann-Godo}, {Nolan}, {Norris}, {Nuss}, {Ohsugi},
  {Okumura}, {Omodei}, {Orlando}, {Ormes}, {Ozaki}, {Paneque}, {Panetta},
  {Parent}, {Pelassa}, {Pepe}, {Pesce-Rollins}, {Petrosian}, {Pinchera},
  {Piron}, {Porter}, {Profumo}, {Rain{\`o}}, {Rando}, {Rapposelli}, {Razzano},
  {Reimer}, {Reimer}, {Reposeur}, {Ripken}, {Ritz}, {Rochester}, {Romani},
  {Roth}, {Sadrozinski}, {Saggini}, {Sanchez}, {Sander}, {Sgr{\`o}}, {Siskind},
  {Smith}, {Spandre}, {Spinelli}, {Stawarz}, {Stephens}, {Strickman}, {Strong},
  {Suson}, {Tajima}, {Takahashi}, {Takahashi}, {Tanaka}, {Thayer}, {Thayer},
  {Thompson}, {Tibaldo}, {Tibolla}, {Torres}, {Tosti}, {Tramacere}, {Turri},
  {Uchiyama}, {Usher}, {Vandenbroucke}, {Vasileiou}, {Vilchez}, {Vitale},
  {Waite}, {Wallace}, {Wang}, {Winer}, {Wood}, {Yang}, {Ylinen}, \&
  {Ziegler}}]{Ackermann+2010}
{Ackermann}, M., {Ajello}, M., {Atwood}, W.~B., {et~al.} 2010, \prd, 82, 092004

\bibitem[{{Adriani} {et~al.}(2011){Adriani}, {Barbarino}, {Bazilevskaya},
  {Bellotti}, {Boezio}, {Bogomolov}, {Bongi}, {Bonvicini}, {Borisov}, {Bottai},
  {Bruno}, {Cafagna}, {Campana}, {Carbone}, {Carlson}, {Casolino},
  {Castellini}, {Consiglio}, {de Pascale}, {de Santis}, {de Simone}, {di
  Felice}, {Galper}, {Gillard}, {Grishantseva}, {Jerse}, {Karelin},
  {Koldashov}, {Krutkov}, {Kvashnin}, {Leonov}, {Malakhov}, {Malvezzi},
  {Marcelli}, {Mayorov}, {Menn}, {Mikhailov}, {Mocchiutti}, {Monaco}, {Mori},
  {Nikonov}, {Osteria}, {Palma}, {Papini}, {Pearce}, {Picozza}, {Pizzolotto},
  {Ricci}, {Ricciarini}, {Rossetto}, {Sarkar}, {Simon}, {Sparvoli},
  {Spillantini}, {Stochaj}, {Stockton}, {Stozhkov}, {Vacchi}, {Vannuccini},
  {Vasilyev}, {Voronov}, {Wu}, {Yurkin}, {Zampa}, {Zampa}, \&
  {Zverev}}]{Adriani+2011}
{Adriani}, O., {Barbarino}, G.~C., {Bazilevskaya}, G.~A., {et~al.} 2011, \prl,
  106, 201101

\bibitem[{{Aguilar} {et~al.}(2014){Aguilar}, {Aisa}, {Alvino}, {Ambrosi},
  {Andeen}, {Arruda}, {Attig}, {Azzarello}, {Bachlechner}, {Barao}, {Barrau},
  {Barrin}, {Bartoloni}, {Basara}, {Battarbee}, {Battiston}, {Bazo}, {Becker},
  {Behlmann}, {Beischer}, {Berdugo}, {Bertucci}, {Bigongiari}, {Bindi},
  {Bizzaglia}, {Bizzarri}, {Boella}, {de Boer}, {Bollweg}, {Bonnivard},
  {Borgia}, {Borsini}, {Boschini}, {Bourquin}, {Burger}, {Cadoux}, {Cai},
  {Capell}, {Caroff}, {Casaus}, {Cascioli}, {Castellini}, {Cernuda},
  {Cervelli}, {Chae}, {Chang}, {Chen}, {Chen}, {Cheng}, {Chen}, {Cheng},
  {Chikanian}, {Chou}, {Choumilov}, {Choutko}, {Chung}, {Clark}, {Clavero},
  {Coignet}, {Consolandi}, {Contin}, {Corti}, {Coste}, {Cui}, {Dai}, {Delgado},
  {Della Torre}, {Demirk{\"o}z}, {Derome}, {Di Falco}, {Di Masso}, {Dimiccoli},
  {D{\'\i}az}, {von Doetinchem}, {Du}, {Duranti}, {D'Urso}, {Eline}, {Eppling},
  {Eronen}, {Fan}, {Farnesini}, {Feng}, {Fiandrini}, {Fiasson}, {Finch},
  {Fisher}, {Galaktionov}, {Gallucci}, {Garc{\'\i}a}, {Garc{\'\i}a-L{\'o}pez},
  {Gast}, {Gebauer}, {Gervasi}, {Ghelfi}, {Gillard}, {Giovacchini}, {Goglov},
  {Gong}, {Goy}, {Grabski}, {Grand i}, {Graziani}, {Guandalini}, {Guerri},
  {Guo}, {Habiby}, {Haino}, {Han}, {He}, {Heil}, {Hoffman}, {Hsieh}, {Huang},
  {Huh}, {Incagli}, {Ionica}, {Jang}, {Jinchi}, {Kanishev}, {Kim}, {Kim},
  {Kirn}, {Kossakowski}, {Kounina}, {Kounine}, {Koutsenko}, {Krafczyk}, {Kunz},
  {La Vacca}, {Laudi}, {Laurenti}, {Lazzizzera}, {Lebedev}, {Lee}, {Lee},
  {Leluc}, {Li}, {Li}, {Li}, {Li}, {Li}, {Li}, {Li}, {Li}, {Li}, {Lim}, {Lin},
  {Lipari}, {Lippert}, {Liu}, {Liu}, {Lomtadze}, {Lu}, {Lu}, {Luebelsmeyer},
  {Luo}, {Luo}, {Lv}, {Majka}, {Malinin}, {Ma{\~n}{\'a}}, {Mar{\'\i}n},
  {Martin}, {Mart{\'\i}nez}, {Masi}, {Maurin}, {Menchaca-Rocha}, {Meng}, {Mo},
  {Morescalchi}, {Mott}, {M{\"u}ller}, {Ni}, {Nikonov}, {Nozzoli}, {Nunes},
  {Obermeier}, {Oliva}, {Orcinha}, {Palmonari}, {Palomares}, {Paniccia},
  {Papi}, {Pedreschi}, {Pensotti}, {Pereira}, {Pilo}, {Piluso}, {Pizzolotto},
  {Plyaskin}, {Pohl}, {Poireau}, {Postaci}, {Putze}, {Quadrani}, {Qi},
  {Rancoita}, {Rapin}, {Ricol}, {Rodr{\'\i}guez}, {Rosier-Lees}, {Rozhkov},
  {Rozza}, {Sagdeev}, {Sandweiss}, {Saouter}, {Sbarra}, {Schael}, {Schmidt},
  {Schuckardt}, {von Dratzig}, {Schwering}, {Scolieri}, {Seo}, {Shan}, {Shan},
  {Shi}, {Shi}, {Shi}, {Siedenburg}, {Son}, {Spada}, {Spinella}, {Sun}, {Sun},
  {Tacconi}, {Tang}, {Tang}, {Tang}, {Tao}, {Tescaro}, {Ting}, {Ting},
  {Tomassetti}, {Torsti}, {T{\"u}rko{\v{g}}lu}, {Urban}, {Vagelli}, {Valente},
  {Vannini}, {Valtonen}, {Vaurynovich}, {Vecchi}, {Velasco}, {Vialle}, {Wang},
  {Wang}, {Wang}, {Wang}, {Wang}, {Weng}, {Whitman}, {Wienkenh{\"o}ver}, {Wu},
  {Xia}, {Xie}, {Xie}, {Xiong}, {Xin}, {Xu}, {Xu}, {Yan}, {Yang}, {Yang}, {Ye},
  {Yi}, {Yu}, {Yu}, {Zeissler}, {Zhang}, {Zhang}, {Zhang}, {Zhang}, {Zheng},
  {Zhuang}, {Zhukov}, {Zichichi}, {Zimmermann}, {Zuccon}, {Zurbach}, \& {AMS
  Collaboration}}]{Aguilar+2014}
{Aguilar}, M., {Aisa}, D., {Alvino}, A., {et~al.} 2014, \prl, 113, 121102

\bibitem[{{Alvarez} {et~al.}(1997){Alvarez}, {Aparici}, {May}, \&
  {Olmos}}]{Alvarez+1997}
{Alvarez}, H., {Aparici}, J., {May}, J., \& {Olmos}, F. 1997, \aaps, 124, 205

\bibitem[{{Beck}(2015)}]{Beck2015}
{Beck}, R. 2015, \aapr, 24, 4

\bibitem[{{Bellomi} {et~al.}(2020){Bellomi}, {Godard}, {Hennebelle},
  {Valdivia}, {Pineau des For{\^e}ts}, {Lesaffre}, \&
  {P{\'e}rault}}]{Bellomi+2020}
{Bellomi}, E., {Godard}, B., {Hennebelle}, P., {et~al.} 2020, \aap, 643, A36

\bibitem[{{Berger} \& {Oliger}(1984)}]{berger_adaptive_1984}
{Berger}, M.~J. \& {Oliger}, J. 1984, Journal of Computational Physics, 53, 484

\bibitem[{{Bowman} {et~al.}(2018){Bowman}, {Rogers}, {Monsalve}, {Mozdzen}, \&
  {Mahesh}}]{Bowman2018}
{Bowman}, J.~D., {Rogers}, A. E.~E., {Monsalve}, R.~A., {Mozdzen}, T.~J., \&
  {Mahesh}, N. 2018, \nat, 555, 67

\bibitem[{{Bracco} {et~al.}(2020){Bracco}, {Jeli{\'c}}, {Marchal}, {Turi{\'c}},
  {Erceg}, {Miville-Desch{\^e}nes}, \& {Bellomi}}]{Bracco+2020}
{Bracco}, A., {Jeli{\'c}}, V., {Marchal}, A., {et~al.} 2020, \aap, 644, L3

\bibitem[{Chapman \& Jeli{\'c}(2019)}]{ChapmanJelic19}
Chapman, E. \& Jeli{\'c}, V. 2019, in The Cosmic 21-cm Revolution, 2514-3433
  (IOP Publishing), 6--1 to 6--29

\bibitem[{{Cummings} {et~al.}(2016){Cummings}, {Stone}, {Heikkila}, {Lal},
  {Webber}, {J{\'o}hannesson}, {Moskalenko}, {Orlando}, \&
  {Porter}}]{Cummings+2016}
{Cummings}, A.~C., {Stone}, E.~C., {Heikkila}, B.~C., {et~al.} 2016, \apj, 831,
  18

\bibitem[{{Dewdney} {et~al.}(2009){Dewdney}, {Hall}, {Schilizzi}, \&
  {Lazio}}]{Dewdney+2009}
{Dewdney}, P.~E., {Hall}, P.~J., {Schilizzi}, R.~T., \& {Lazio}, T.~J.~L.~W.
  2009, IEEE Proceedings, 97, 1482

\bibitem[{Federrath {et~al.}(2010)Federrath, Roman-Duval, Klessen, Schmidt, \&
  Mac~Low}]{federrath_comparing_2010}
Federrath, C., Roman-Duval, J., Klessen, R.~S., Schmidt, W., \& Mac~Low, M.-M.
  2010, \aap, 512, A81

\bibitem[{{Ferri{\`e}re}(2020)}]{Ferriere2020}
{Ferri{\`e}re}, K. 2020, Plasma Physics and Controlled Fusion, 62, 014014

\bibitem[{Field(1965)}]{field_thermal_1965}
Field, G.~B. 1965, \apj, 142, 531

\bibitem[{{Field} {et~al.}(1969){Field}, {Goldsmith}, \&
  {Habing}}]{field_cosmic_1969}
{Field}, G.~B., {Goldsmith}, D.~W., \& {Habing}, H.~J. 1969, \apjl, 155, L149

\bibitem[{{Fromang} {et~al.}(2006){Fromang}, {Hennebelle}, \&
  {Teyssier}}]{Fromang2006}
{Fromang}, S., {Hennebelle}, P., \& {Teyssier}, R. 2006, \aap, 457, 371

\bibitem[{{Gaensler} {et~al.}(2011){Gaensler}, {Haverkorn}, {Burkhart},
  {Newton-McGee}, {Ekers}, {Lazarian}, {McClure-Griffiths}, {Robishaw},
  {Dickey}, \& {Green}}]{Gaensler2011}
{Gaensler}, B.~M., {Haverkorn}, M., {Burkhart}, B., {et~al.} 2011, \nat, 478,
  214

\bibitem[{{Gehlot} {et~al.}(2019){Gehlot}, {Mertens}, {Koopmans}, {Brentjens},
  {Zaroubi}, {Ciardi}, {Ghosh}, {Hatef}, {Iliev}, {Jeli{\'c}}, {}, {Kooistra},
  {Krause}, {Mellema}, {Mevius}, {Mitra}, {Offringa}, {Pandey}, {Sardarabadi},
  {Schaye}, {Silva}, {Vedantham}, \& {Yatawatta}}]{Gehlot2019}
{Gehlot}, B.~K., {Mertens}, F.~G., {Koopmans}, L.~V.~E., {et~al.} 2019, \mnras,
  488, 4271

\bibitem[{{Ginzburg} \& {Syrovatskii}(1964)}]{GinzburgSyrovatskii1964}
{Ginzburg}, V.~L. \& {Syrovatskii}, S.~I. 1964, {The Origin of Cosmic Rays}

\bibitem[{{Ginzburg} \& {Syrovatskii}(1965)}]{GinzburgSyrovatskii1965}
{Ginzburg}, V.~L. \& {Syrovatskii}, S.~I. 1965, \araa, 3, 297

\bibitem[{{Grenier} {et~al.}(2015){Grenier}, {Black}, \&
  {Strong}}]{Grenier+2015}
{Grenier}, I.~A., {Black}, J.~H., \& {Strong}, A.~W. 2015, \araa, 53, 199

\bibitem[{{Guzm{\'a}n} {et~al.}(2011){Guzm{\'a}n}, {May}, {Alvarez}, \&
  {Maeda}}]{Guzman+2011}
{Guzm{\'a}n}, A.~E., {May}, J., {Alvarez}, H., \& {Maeda}, K. 2011, \aap, 525,
  A138

\bibitem[{{Habing}(1968)}]{Habing1968}
{Habing}, H.~J. 1968, \bain, 19, 421

\bibitem[{{Haslam} {et~al.}(1981){Haslam}, {Klein}, {Salter}, {Stoffel},
  {Wilson}, {Cleary}, {Cooke}, \& {Thomasson}}]{Haslam+1981}
{Haslam}, C.~G.~T., {Klein}, U., {Salter}, C.~J., {et~al.} 1981, \aap, 100, 209

\bibitem[{{Haslam} {et~al.}(1982){Haslam}, {Salter}, {Stoffel}, \&
  {Wilson}}]{Haslam+1982}
{Haslam}, C.~G.~T., {Salter}, C.~J., {Stoffel}, H., \& {Wilson}, W.~E. 1982,
  \aaps, 47, 1

\bibitem[{{Haverkorn} {et~al.}(2004){Haverkorn}, {Katgert}, \& {de
  Bruyn}}]{Haverkorn2004}
{Haverkorn}, M., {Katgert}, P., \& {de Bruyn}, A.~G. 2004, \aap, 427, 549

\bibitem[{Heiles \& Troland(2003{\natexlab{a}})}]{heiles_millennium_2003}
Heiles, C. \& Troland, T.~H. 2003{\natexlab{a}}, ApJ, 586, 1067

\bibitem[{Heiles \& Troland(2003{\natexlab{b}})}]{heiles_vizier_2003}
Heiles, C. \& Troland, T.~H. 2003{\natexlab{b}}, {VizieR} {Online} {Data}
  {Catalog}: {Millennium} {Arecibo} 21-cm {Survey} ({Heiles}+, 2003) -
  {NASA}/{ADS}

\bibitem[{{Heiles} \& {Troland}(2005)}]{Heiles2005a}
{Heiles}, C. \& {Troland}, T.~H. 2005, \apj, 624, 773

\bibitem[{{Ivlev} {et~al.}(2015){Ivlev}, {Padovani}, {Galli}, \&
  {Caselli}}]{Ivlev+2015}
{Ivlev}, A.~V., {Padovani}, M., {Galli}, D., \& {Caselli}, P. 2015, \apj, 812,
  135

\bibitem[{{Jeli{\'c}} {et~al.}(2014){Jeli{\'c}}, {de Bruyn}, {Mevius},
  {Abdalla}, {Asad}, {Bernardi}, {Brentjens}, {Bus}, {Chapman}, {Ciardi},
  {Daiboo}, {Fernandez}, {Ghosh}, {Harker}, {Jensen}, {Kazemi}, {Koopmans},
  {Labropoulos}, {Martinez-Rubi}, {Mellema}, {Offringa}, {Pandey}, {Patil},
  {Thomas}, {Vedantham}, {Veligatla}, {Yatawatta}, {Zaroubi}, {Alexov},
  {Anderson}, {Avruch}, {Beck}, {Bell}, {Bentum}, {Best}, {Bonafede},
  {Bregman}, {Breitling}, {Broderick}, {Brouw}, {Br{\"u}ggen}, {Butcher},
  {Conway}, {de Gasperin}, {de Geus}, {Deller}, {Dettmar}, {Duscha},
  {Eisl{\"o}ffel}, {Engels}, {Falcke}, {Fallows}, {Fender}, {Ferrari},
  {Frieswijk}, {Garrett}, {Grie{\ss}meier}, {Gunst}, {Hamaker}, {Hassall},
  {Haverkorn}, {Heald}, {Hessels}, {Hoeft}, {H{\"o}randel}, {Horneffer}, {van
  der Horst}, {Iacobelli}, {Juette}, {Karastergiou}, {Kondratiev}, {Kramer},
  {Kuniyoshi}, {Kuper}, {van Leeuwen}, {Maat}, {Mann}, {McKay-Bukowski},
  {McKean}, {Munk}, {Nelles}, {Norden}, {Paas}, {Pandey-Pommier}, {Pietka},
  {Pizzo}, {Polatidis}, {Reich}, {R{\"o}ttgering}, {Rowlinson}, {Scaife},
  {Schwarz}, {Serylak}, {Smirnov}, {Steinmetz}, {Stewart}, {Tagger}, {Tang},
  {Tasse}, {ter Veen}, {Thoudam}, {Toribio}, {Vermeulen}, {Vocks}, {van
  Weeren}, {Wijers}, {Wijnholds}, {Wucknitz}, \& {Zarka}}]{Jelic+2014}
{Jeli{\'c}}, V., {de Bruyn}, A.~G., {Mevius}, M., {et~al.} 2014, \aap, 568,
  A101

\bibitem[{{Jeli{\'c}} {et~al.}(2015){Jeli{\'c}}, {de Bruyn}, {Pandey},
  {Mevius}, {Haverkorn}, {Brentjens}, {Koopmans}, {Zaroubi}, {Abdalla}, {Asad},
  {Bus}, {Chapman}, {Ciardi}, {Fernandez}, {Ghosh}, {Harker}, {Iliev},
  {Jensen}, {Kazemi}, {Mellema}, {Offringa}, {Patil}, {Vedantham}, \&
  {Yatawatta}}]{Jelic+2015}
{Jeli{\'c}}, V., {de Bruyn}, A.~G., {Pandey}, V.~N., {et~al.} 2015, \aap, 583,
  A137

\bibitem[{{Kalberla} \& {Haud}(2018)}]{kalberla_properties_2018}
{Kalberla}, P.~M.~W. \& {Haud}, U. 2018, \aap, 619, A58

\bibitem[{Khokhlov(1998)}]{khokhlov_fully_1998}
Khokhlov, A.~M. 1998, Journal of Computational Physics, 143, 519

\bibitem[{{Longair}(2011)}]{Longair2011}
{Longair}, M.~S. 2011, {High Energy Astrophysics}

\bibitem[{{Maeda} {et~al.}(1999){Maeda}, {Alvarez}, {Aparici}, {May}, \&
  {Reich}}]{Maeda+1999}
{Maeda}, K., {Alvarez}, H., {Aparici}, J., {May}, J., \& {Reich}, P. 1999,
  \aaps, 140, 145

\bibitem[{{Marchal} {et~al.}(2019){Marchal}, {Miville-Desch{\^e}nes}, {Orieux},
  {Gac}, {Soussen}, {Lesot}, {d'Allonnes}, \&
  {Salom{\'e}}}]{marchal_rohsa_2019}
{Marchal}, A., {Miville-Desch{\^e}nes}, M.-A., {Orieux}, F., {et~al.} 2019,
  A\&A, 626, A101

\bibitem[{{Mertens} {et~al.}(2020){Mertens}, {Mevius}, {Koopmans}, {Offringa},
  {Mellema}, {Zaroubi}, {Brentjens}, {Gan}, {Gehlot}, {Pandey}, {Sardarabadi},
  {Vedantham}, {Yatawatta}, {Asad}, {Ciardi}, {Chapman}, {Gazagnes}, {Ghara},
  {Ghosh}, {Giri}, {Iliev}, {Jeli{\'c}}, {Kooistra}, {Mondal}, {Schaye}, \&
  {Silva}}]{Mertens2020}
{Mertens}, F.~G., {Mevius}, M., {Koopmans}, L.~V.~E., {et~al.} 2020, \mnras,
  493, 1662

\bibitem[{{Mozdzen} {et~al.}(2017){Mozdzen}, {Bowman}, {Monsalve}, \&
  {Rogers}}]{Mozdzen2017}
{Mozdzen}, T.~J., {Bowman}, J.~D., {Monsalve}, R.~A., \& {Rogers}, A.~E.~E.
  2017, \mnras, 464, 4995

\bibitem[{{Mozdzen} {et~al.}(2019){Mozdzen}, {Mahesh}, {Monsalve}, {Rogers}, \&
  {Bowman}}]{Mozdzen2019}
{Mozdzen}, T.~J., {Mahesh}, N., {Monsalve}, R.~A., {Rogers}, A.~E.~E., \&
  {Bowman}, J.~D. 2019, \mnras, 483, 4411

\bibitem[{Murray {et~al.}(2015)Murray, Stanimirovi{\'c}, Goss, Dickey, Heiles,
  Lindner, Babler, Pingel, Lawrence, Jencson, \&
  Hennebelle}]{murray_21-sponge_2015}
Murray, C.~E., Stanimirovi{\'c}, S., Goss, W.~M., {et~al.} 2015, ApJ, 804, 89

\bibitem[{Murray {et~al.}(2018)Murray, Stanimirovi{\'c}, Goss, Heiles, Dickey,
  Babler, \& Kim}]{murray_21-sponge_2018}
Murray, C.~E., Stanimirovi{\'c}, S., Goss, W.~M., {et~al.} 2018, ApJS, 238, 14

\bibitem[{{Orlando}(2018)}]{Orlando2018}
{Orlando}, E. 2018, \mnras, 475, 2724

\bibitem[{{Padovani} {et~al.}(2009){Padovani}, {Galli}, \&
  {Glassgold}}]{Padovani+2009}
{Padovani}, M., {Galli}, D., \& {Glassgold}, A.~E. 2009, \aap, 501, 619

\bibitem[{{Padovani} {et~al.}(2018){Padovani}, {Ivlev}, {Galli}, \&
  {Caselli}}]{Padovani+2018a}
{Padovani}, M., {Ivlev}, A.~V., {Galli}, D., \& {Caselli}, P. 2018, \aap, 614,
  A111

\bibitem[{{Planck Collaboration Int. XX}(2015)}]{PlanckXX2015}
{Planck Collaboration Int. XX}. 2015, \aap, 576, A105

\bibitem[{{Planck Collaboration IV}(2020)}]{Planck2018IV}
{Planck Collaboration IV}. 2020, \aap, 641, A4

\bibitem[{{Platania} {et~al.}(1998){Platania}, {Bensadoun}, {Bersanelli}, {De
  Amici}, {Kogut}, {Levin}, {Maino}, \& {Smoot}}]{Platania+1998}
{Platania}, P., {Bensadoun}, M., {Bersanelli}, M., {et~al.} 1998, \apj, 505,
  473

\bibitem[{{Reich} \& {Reich}(1988{\natexlab{a}})}]{ReichReich88a}
{Reich}, P. \& {Reich}, W. 1988{\natexlab{a}}, \aaps, 74, 7

\bibitem[{{Reich} \& {Reich}(1988{\natexlab{b}})}]{ReichReich88b}
{Reich}, P. \& {Reich}, W. 1988{\natexlab{b}}, \aap, 196, 211

\bibitem[{{Reissl} {et~al.}(2019){Reissl}, {Brauer}, {Klessen}, \&
  {Pellegrini}}]{Reissl+2019}
{Reissl}, S., {Brauer}, R., {Klessen}, R.~S., \& {Pellegrini}, E.~W. 2019,
  \apj, 885, 15

\bibitem[{{Roger} {et~al.}(1999){Roger}, {Costain}, {Landecker}, \&
  {Swerdlyk}}]{Roger+1999}
{Roger}, R.~S., {Costain}, C.~H., {Landecker}, T.~L., \& {Swerdlyk}, C.~M.
  1999, \aaps, 137, 7

\bibitem[{{Rybicki} \& {Lightman}(1986)}]{RybickiLightman86}
{Rybicki}, G.~B. \& {Lightman}, A.~P. 1986, {Radiative Processes in
  Astrophysics}, 400

\bibitem[{Schmidt {et~al.}(2009)Schmidt, Federrath, Hupp, Kern, \&
  Niemeyer}]{schmidt_numerical_2009}
Schmidt, W., Federrath, C., Hupp, M., Kern, S., \& Niemeyer, J.~C. 2009, \aap,
  494, 127

\bibitem[{{Sokoloff} {et~al.}(1998){Sokoloff}, {Bykov}, {Shukurov},
  {Berkhuijsen}, {Beck}, \& {Poezd}}]{Sokoloff1998}
{Sokoloff}, D.~D., {Bykov}, A.~A., {Shukurov}, A., {et~al.} 1998, \mnras, 299,
  189

\bibitem[{{Stone} {et~al.}(2019){Stone}, {Cummings}, {Heikkila}, \&
  {Lal}}]{Stone+2019}
{Stone}, E.~C., {Cummings}, A.~C., {Heikkila}, B.~C., \& {Lal}, N. 2019, Nature
  Astronomy, 3, 1013

\bibitem[{{Strong} {et~al.}(2007){Strong}, {Moskalenko}, \&
  {Ptuskin}}]{Strong+2007}
{Strong}, A.~W., {Moskalenko}, I.~V., \& {Ptuskin}, V.~S. 2007, Annual Review
  of Nuclear and Particle Science, 57, 285

\bibitem[{{Sun} {et~al.}(2008){Sun}, {Reich}, {Waelkens}, \&
  {En{\ss}lin}}]{Sun+2008}
{Sun}, X.~H., {Reich}, W., {Waelkens}, A., \& {En{\ss}lin}, T.~A. 2008, \aap,
  477, 573

\bibitem[{{Teyssier}(2002)}]{Teyssier2002}
{Teyssier}, R. 2002, \aap, 385, 337

\bibitem[{{Trott} {et~al.}(2020){Trott}, {Jordan}, {Midgley}, {Barry}, {Greig},
  {Pindor}, {Cook}, {Sleap}, {Tingay}, {Ung}, {Hancock}, {Williams}, {Bowman},
  {Byrne}, {Chokshi}, {Hazelton}, {Hasegawa}, {Jacobs}, {Joseph}, {Li}, {Line},
  {Lynch}, {McKinley}, {Mitchell}, {Morales}, {Ouchi}, {Pober}, {Rahimi},
  {Takahashi}, {Wayth}, {Webster}, {Wilensky}, {Wyithe}, {Yoshiura}, {Zhang},
  \& {Zheng}}]{Trott2020}
{Trott}, C.~M., {Jordan}, C.~H., {Midgley}, S., {et~al.} 2020, \mnras, 493,
  4711

\bibitem[{{Van Eck} {et~al.}(2017){Van Eck}, {Haverkorn}, {Alves}, {Beck}, {de
  Bruyn}, {En{\ss}lin}, {Farnes}, {Ferri{\`e}re}, {Heald}, {Horellou},
  {Horneffer}, {Iacobelli}, {Jeli{\'c}}, {Mart{\'\i}-Vidal}, {Mulcahy},
  {Reich}, {R{\"o}ttgering}, {Scaife}, {Schnitzeler}, {Sobey}, \&
  {Sridhar}}]{vanEck2017}
{Van Eck}, C.~L., {Haverkorn}, M., {Alves}, M.~I.~R., {et~al.} 2017, \aap, 597,
  A98

\bibitem[{{van Haarlem} {et~al.}(2013){van Haarlem}, {Wise}, {Gunst}, {Heald},
  {McKean}, {Hessels}, {de Bruyn}, {Nijboer}, {Swinbank}, {Fallows},
  {Brentjens}, {Nelles}, {Beck}, {Falcke}, {Fender}, {H{\"o}randel},
  {Koopmans}, {Mann}, {Miley}, {R{\"o}ttgering}, {Stappers}, {Wijers},
  {Zaroubi}, {van den Akker}, {Alexov}, {Anderson}, {Anderson}, {van Ardenne},
  {Arts}, {Asgekar}, {Avruch}, {Batejat}, {B{\"a}hren}, {Bell}, {Bell}, {van
  Bemmel}, {Bennema}, {Bentum}, {Bernardi}, {Best}, {B{\^\i}rzan}, {Bonafede},
  {Boonstra}, {Braun}, {Bregman}, {Breitling}, {van de Brink}, {Broderick},
  {Broekema}, {Brouw}, {Br{\"u}ggen}, {Butcher}, {van Cappellen}, {Ciardi},
  {Coenen}, {Conway}, {Coolen}, {Corstanje}, {Damstra}, {Davies}, {Deller},
  {Dettmar}, {van Diepen}, {Dijkstra}, {Donker}, {Doorduin}, {Dromer}, {Drost},
  {van Duin}, {Eisl{\"o}ffel}, {van Enst}, {Ferrari}, {Frieswijk}, {Gankema},
  {Garrett}, {de Gasperin}, {Gerbers}, {de Geus}, {Grie{\ss}meier}, {Grit},
  {Gruppen}, {Hamaker}, {Hassall}, {Hoeft}, {Holties}, {Horneffer}, {van der
  Horst}, {van Houwelingen}, {Huijgen}, {Iacobelli}, {Intema}, {Jackson},
  {Jelic}, {de Jong}, {Juette}, {Kant}, {Karastergiou}, {Koers}, {Kollen},
  {Kondratiev}, {Kooistra}, {Koopman}, {Koster}, {Kuniyoshi}, {Kramer},
  {Kuper}, {Lambropoulos}, {Law}, {van Leeuwen}, {Lemaitre}, {Loose}, {Maat},
  {Macario}, {Markoff}, {Masters}, {McFadden}, {McKay-Bukowski}, {Meijering},
  {Meulman}, {Mevius}, {Middelberg}, {Millenaar}, {Miller-Jones}, {Mohan},
  {Mol}, {Morawietz}, {Morganti}, {Mulcahy}, {Mulder}, {Munk}, {Nieuwenhuis},
  {van Nieuwpoort}, {Noordam}, {Norden}, {Noutsos}, {Offringa}, {Olofsson},
  {Omar}, {Orr{\'u}}, {Overeem}, {Paas}, {Pand ey-Pommier}, {Pandey}, {Pizzo},
  {Polatidis}, {Rafferty}, {Rawlings}, {Reich}, {de Reijer}, {Reitsma},
  {Renting}, {Riemers}, {Rol}, {Romein}, {Roosjen}, {Ruiter}, {Scaife}, {van
  der Schaaf}, {Scheers}, {Schellart}, {Schoenmakers}, {Schoonderbeek},
  {Serylak}, {Shulevski}, {Sluman}, {Smirnov}, {Sobey}, {Spreeuw}, {Steinmetz},
  {Sterks}, {Stiepel}, {Stuurwold}, {Tagger}, {Tang}, {Tasse}, {Thomas},
  {Thoudam}, {Toribio}, {van der Tol}, {Usov}, {van Veelen}, {van der Veen},
  {ter Veen}, {Verbiest}, {Vermeulen}, {Vermaas}, {Vocks}, {Vogt}, {de Vos},
  {van der Wal}, {van Weeren}, {Weggemans}, {Weltevrede}, {White}, {Wijnholds},
  {Wilhelmsson}, {Wucknitz}, {Yatawatta}, {Zarka}, {Zensus}, \& {van
  Zwieten}}]{vanHaarlem2013}
{van Haarlem}, M.~P., {Wise}, M.~W., {Gunst}, A.~W., {et~al.} 2013, \aap, 556,
  A2

\bibitem[{{Waelkens} {et~al.}(2009){Waelkens}, {Jaffe}, {Reinecke}, {Kitaura},
  \& {En{\ss}lin}}]{Waelkens+2009}
{Waelkens}, A., {Jaffe}, T., {Reinecke}, M., {Kitaura}, F.~S., \& {En{\ss}lin},
  T.~A. 2009, \aap, 495, 697

\bibitem[{{Wang} {et~al.}(2020){Wang}, {Jaffe}, {En{\ss}lin}, {Ullio}, {Ghosh},
  \& {Santos}}]{Wang+2020}
{Wang}, J., {Jaffe}, T.~R., {En{\ss}lin}, T.~A., {et~al.} 2020, \apjs, 247, 18

\bibitem[{{Wolfire} {et~al.}(2003){Wolfire}, {McKee}, {Hollenbach}, \&
  {Tielens}}]{Wolfire2003}
{Wolfire}, M.~G., {McKee}, C.~F., {Hollenbach}, D., \& {Tielens}, A.~G.~G.~M.
  2003, \apj, 587, 278

\end{thebibliography}

\appendix

%
%

\section{Bivariate distributions of $\mu$ and $\beta$ at high frequencies}
\label{app:betaLMH}

\begin{figure}[!h]
\begin{center}
\resizebox{1\hsize}{!}{\includegraphics{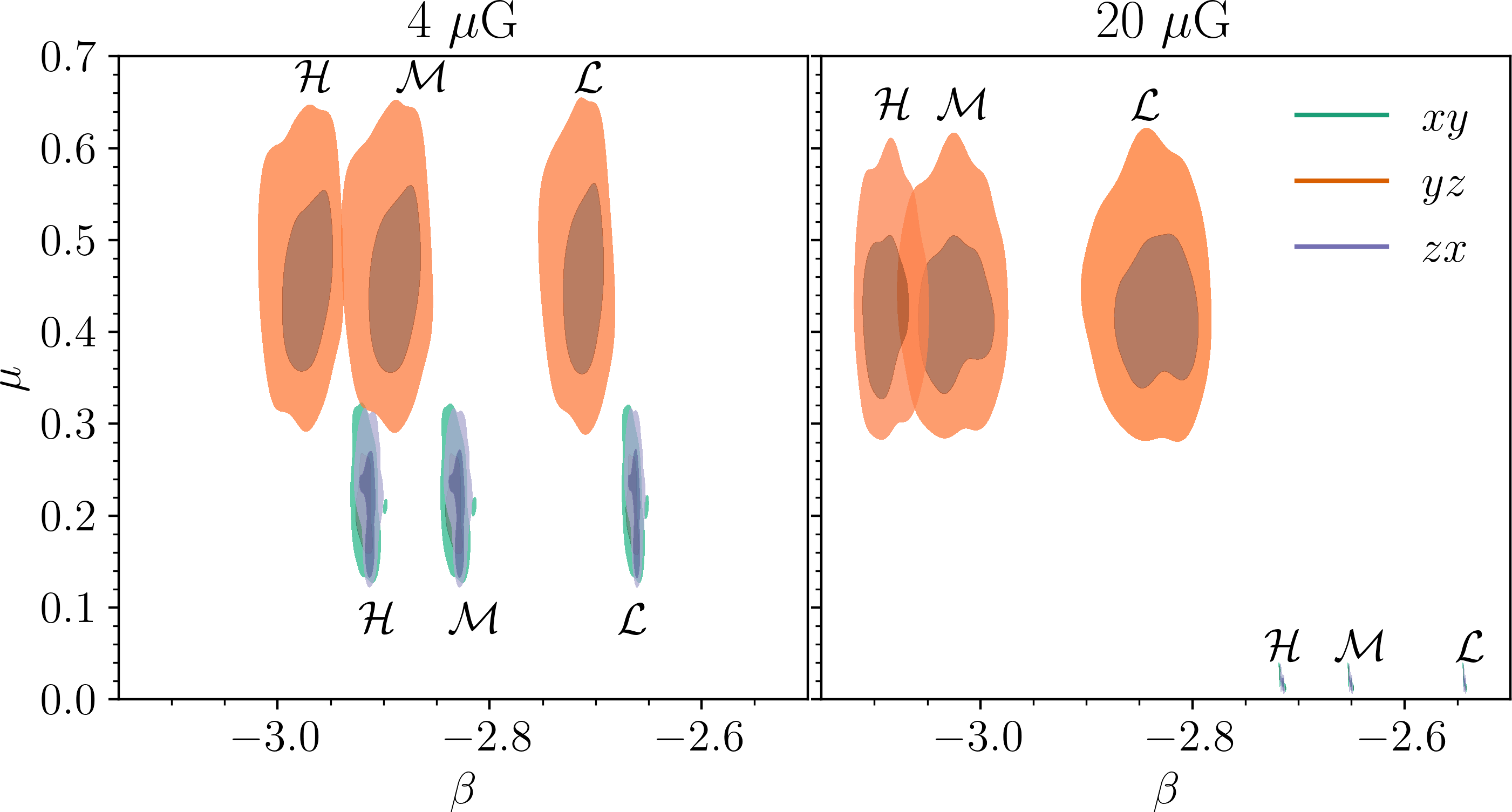}}
\caption{Bivariate distribution of the ratio between the standard deviation of \bperp\ and
its median value, $\mu$, and the brightness temperature spectral index, $\beta$,
computed in the frequency ranges 115$-$189~MHz, 467$-$672~MHz, and 833$-$1200~MHz
(labelled $\mathcal{L}$, $\mathcal{M}$, and $\mathcal{H}$, respectively) 
for each LOS for the weak and strong field case (left and right panel, respectively),
at a resolution of $6.7'$.
The POS are identified by the three different colours displayed in the legend. Isodensity contours
are plotted at 50\% and 80\%.
}
\label{betamu3freq}
\end{center}
\end{figure}

Following the procedure outlined in Sect.~\ref{ssec:beta}, we consider two higher frequency ranges, 
467$-$672~MHz and 833$-$1200~MHz (labelled $\mathcal{M}$ and $\mathcal{H}$, respectively), 
for the calculation of $\beta$ in the three POS of the two simulation snapshots. 
Figure~\ref{betamu3freq} shows the bivariate distribution of $\mu$ and $\beta$ for these two 
frequency ranges, comparing them with the distribution for the range 115$-$189~MHz 
(labelled $\mathcal{L}$) described in Sect.~\ref{ssec:beta}.
In order to increase clarity and to avoid overlapping distributions, 
we only show density isodensity contours plotted at 50\% and 80\%. 
As the frequency range increases, the $\beta$ distributions shift towards more negative values. 
This can be explained by looking at Fig.~\ref{depsdE_and_s_vs_E} where we see that as the frequency increases, 
the energies determining the synchrotron emissivity are higher and higher. 
Higher energies correspond to more negative values of $s$, and therefore of $\beta$.
For completeness, Table~\ref{tab:betaLMH} 
shows the values of $\tilde\beta$ 
for the three POS of the
two snapshots considered in the frequency intervals $\mathcal{M}$ and $\mathcal{H}$.

\begin{table}[!h]
    \caption{ 
    Medians of 
    the brightness temperature spectral index
    for the three POS and the two cases of weak and strong field at a resolution of $6.7'$. 
	Labels $\mathcal{M}$ and $\mathcal{H}$ identify the frequency intervals
	467$-$672~MHz and 833$-$1200~MHz, respectively$^{a}$.}

\begin{center}
\resizebox{\linewidth}{!}{
\begin{tabular}{cccc}
\toprule\toprule
&POS & \multicolumn{2}{c}{$\tilde\beta$}\\
\cmidrule{3-4}
&& $\mathcal{M}$ & $\mathcal{H}$\\
\midrule
\multirow{3}{*}{\rotatebox[origin=c]{0}{\parbox[c]{2cm}{\centering weak field ($B_{0}=4~\mu$G)}}}
&$xy$ & $-2.83\pm0.01$ & $-2.92\pm0.01$\\
&$yz$ & $-2.89\pm0.02$ & $-2.98\pm0.02$\\
&$zx$ & $-2.83\pm0.01$ & $-2.92\pm0.01$\\
\midrule
\multirow{3}{*}{\rotatebox[origin=c]{0}{\parbox[c]{2cm}{\centering strong field ($B_{0}=20~\mu$G)}}}
&$xy$ & $-2.65$ & $-2.72$\\
&$yz$ & $-3.02\pm0.03$ & $-3.08\pm0.02$\\
&$zx$ & $-2.65$ & $-2.72$\\
\bottomrule
\end{tabular}
}
\end{center}
{\small
\begin{flushleft}
$^a$~~Errors have
	been estimated using the first and third quartiles.
	Errors smaller than 0.01 are not shown.
\end{flushleft}
}
\label{tab:betaLMH}
\end{table}

\section{Brightness temperature spectral index maps}
\label{app:beta}

Here we show the spectral index maps obtained for the two 
simulations described in Sect.~\ref{ssec:sims}. 
These maps have been derived considering the same frequency range 
and frequency resolution of LOFAR observations 
by \citet{Jelic+2015}, namely 
$\nu=115-189$~MHz and $\Delta\nu=183$~kHz, respectively. 
In the POS where 
$\mu\ll1$,
namely the $xy$ and $zx$ POS of 
the strong field case,
$\beta$ shows a constant value. 
Then, the greater the turbulent component of the field, 
the greater the variations that $\beta$ exhibits on small scales. 
The latter can in principle be resolved by observations with LOFAR. 
From these maps, the histograms displayed in 
Fig.~\ref{histobeta} were produced.
\begin{figure}[!h]
\begin{center}
\resizebox{0.9\hsize}{!}{\includegraphics{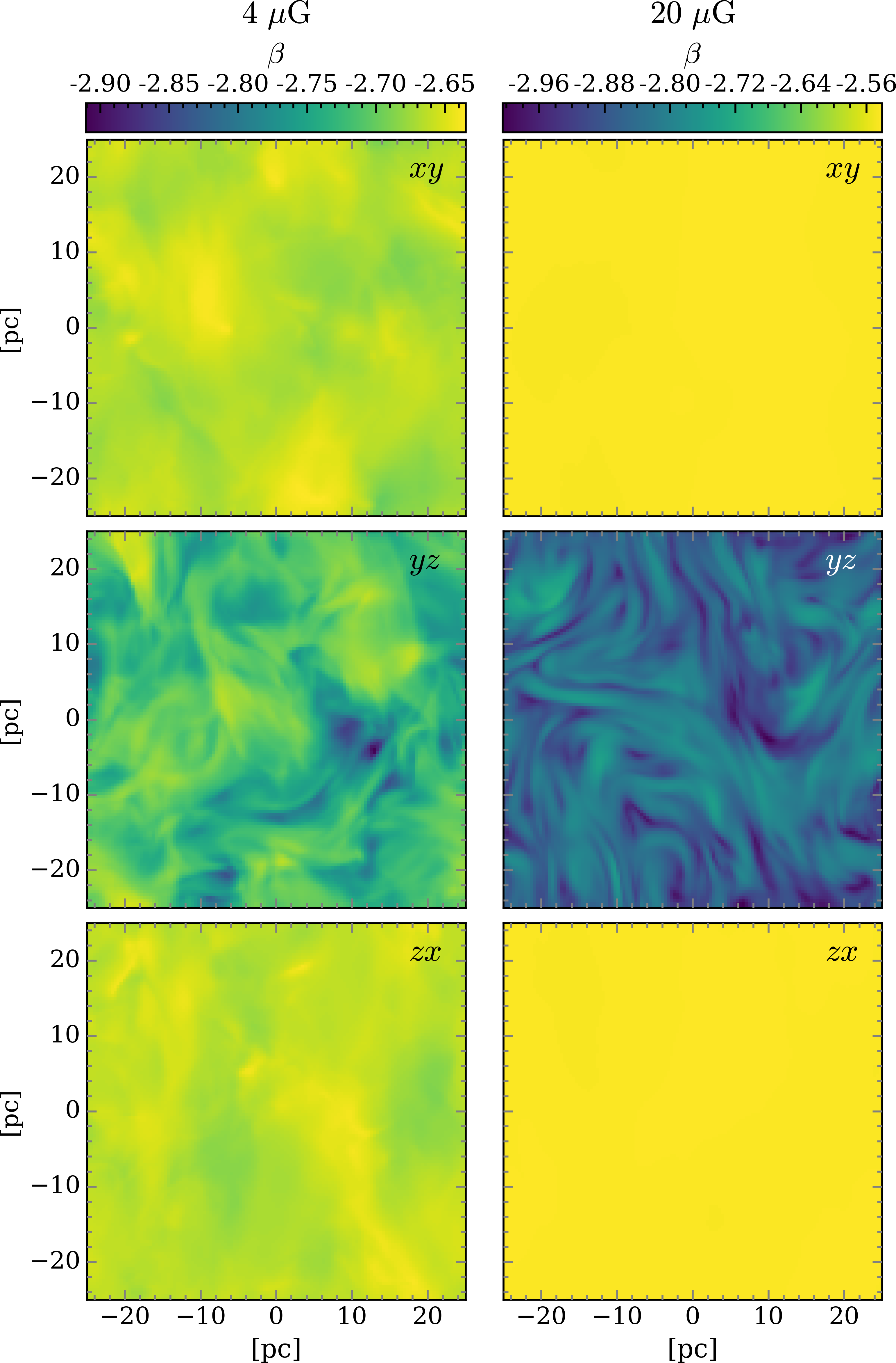}}
\caption{Brightness temperature spectral index maps 
for the three POS (three rows)
of the weak and strong field case
(left and right column, respectively)
for a resolution of $6.7'$.
The three panels of each column share the same colourbar on the top.
}
\label{betamaps420}
\end{center}
\end{figure}

\end{document}